\documentclass[apj,iop]{emulateapj}

\usepackage{amsmath}
\usepackage{tabularx}
\usepackage{natbib}
\usepackage{float}
\usepackage[colorlinks=true,linkcolor=blue,citecolor=blue]{hyperref}%
\usepackage[all]{hypcap}
\usepackage{lineno}
\usepackage{footnote}

\shorttitle{Retrievals on Reflected Light}
\shortauthors{Mukherjee et al.}
\graphicspath{{./}{figures/}}
\begin{document}

\title{Cloud Parameterizations and their Effect on Retrievals of Exoplanet Reflection Spectroscopy}

\email{samukher@ucsc.edu}

\author{Sagnick Mukherjee$^{1,2}$, Natasha E. Batalha$^{3}$, Mark S. Marley$^{3}$}
\affiliation{{$^1$}Department of Astronomy and Astrophysics, UC Santa Cruz, Santa Cruz, CA 95060, USA \\
{$^2$}Department of Physics, Presidency University, 86/1 College Street, Kolkata 700073, India \\
{$^3$}NASA Ames Research Centre, MS 245-3, Moffett Field, CA 94035, USA}
%\affiliation{}

%\affiliation{}

%%\collaboration{1}{(AAS Journals Data Scientists collaboration)}

%%\author{Butler Burton}
%\affiliation{Leiden University}
%\affiliation{AAS Journals Associate Editor-in-Chief}
%\nocollaboration{1}

%\altaffiliation{AASTeX v6+ programmer}
%% check spelling parameterizations
%\collaboration{1}{(LaTeX collaboration)}

%\affiliation{AAS Director of Publishing}
%\affiliation{American Astronomical Society \\
%1667 K Street NW, Suite 800 \\

%\author{Scott Chernoff}
%\affiliation{IOP Publishing, Washington, DC 20005}




\begin{abstract}
Future space-based direct imaging missions will perform low-resolution (R$<$100) optical (0.3-1~$\mu$m) spectroscopy of planets, thus enabling reflected spectroscopy of cool giants. Reflected light spectroscopy is encoded with rich information about the scattering and absorbing properties of planet atmospheres. Given the diversity of clouds and hazes expected in exoplanets, it is imperative we solidify the methodology to accurately and precisely retrieve these scattering and absorbing properties that are agnostic to cloud species. In particular, we focus on determining how different cloud parameterizations affect resultant inferences of both cloud and atmospheric composition. We simulate mock observations of the reflected spectra from three top priority direct imaging cool giant targets with different effective temperatures, ranging from 135 K to 533 K. We perform retrievals of cloud structure and molecular abundances on these three planets using four different parameterizations, each with increasing levels of cloud complexity. We find that the retrieved atmospheric and scattering properties strongly depend on the choice of cloud parameterization. For example, parameterizations that are too simplistic tend to overestimate the abundances. Overall, we are unable to retrieve precise/accurate gravity beyond $\pm$50\%. Lastly, we find that even low SNR=5, low R=40 reflected light spectroscopy gives cursory zeroth order insights into cloud deck position relative to molecular and Rayleigh optical depth level.

\end{abstract}


\section{Introduction} 

Reflected light spectroscopy of exoplanets will be important in the upcoming decades. Future space-based direct imaging missions will perform space-based optical coronagraphy, which will enable low resolution (R $\sim$ 40-200) optical spectra of directly imaged cool giant, and temperate terrestrial planets orbiting Sun-like stars. Although HST/JWST may be able to obtain transmission spectroscopy of a few number of favorable targets (e.g. HIP 41378 f ; \citet{dressing2020HIP}), reflected light offers the opportunity to directly infer the scattering properties of the atmosphere. Reflected light spectra contains key information related to the planet's scattering and chemical properties. Specifically, unlike thermal and transmission spectroscopy, clouds and hazes determine the zeroth order structure of the reflection spectra \citep{nayak17, Lupu_2016, gao2017sulfur, macdonald2018exploring,marley1999reflected}. Given the diversity of clouds and hazes expected in exoplanets \citep{morley14water,Gao20aerosol}, this presents new challenges for retrieving properties from reflected light spectra of exoplanets. 
%This will allow for the in-depth characterization of cloud properties of cool giant planets for the first time. Such cool giant planets will be dominated by reflected starlight in the optical. 

\subsection{Previous Parameterizations used in Reflected Light Retrievals}

In order to prepare for this next decade of exoplanet spectroscopy there has been a growing body of literature aimed at determining best practices in retrieving properties from reflected light observations. We discuss each modeling framework below, focusing on the methodologies for parameterizing clouds. Given our focus on determining the required cloud complexity needed for retrievals, it is important to understand the parameterization method of each of these previous works.

\subsubsection{\citet{Lupu_2016,nayak17}}

\citet{Lupu_2016} showed that the presence/absence of clouds and CH$_4$ can be inferred with high confidence from cool giants with reflection spectroscopy. The forward model used by \citet{Lupu_2016} was based on the model initially developed by \citet{mckay1989thermal, marley1999thermal,marley1999reflected} and later updated by \citet{cahoy2010exoplanet}. They used CH$_4$ molecular opacity and the collision induced opacities of CH$_4$, H$_2$ and He as gaseous opacities in their forward model as they worked only with planets where the reflection spectra is expected to be CH$_4$ dominated. \citet{Lupu_2016} used two simple retrieval models differing in cloud parameterization for retrieving on the reflected spectra, both of which contained wavelength independent clouds. Their first model included a single semi-infinite cloud layer. Their second cloud model had a second cloud deck in addition to the semi-infinite cloud deck. In this case, the semi-infinite cloud deck at the bottom is forced to be optically thick and essentially acts as a reflective surface where the asymmetry parameter for this bottom deck is not retrieved. Using these two models, \citet{Lupu_2016} retrieved on three validation test cases where the simulated data was produced using the retrieval model itself. They also tested their retrieval model on self-consistently modeled test cases of three planets. The performance of their retrieval models showed a decline on the `real' cases compared to the validation cases. For example, only lower limits on the CH$_4$ abundance could be obtained for two of the three real planet cases compared to constraints of CH$_4$ within factors of 10 of the true abundance for retrievals on validation planets. \citet{Lupu_2016} hence note the need to 1) test various other parameterizations with varying levels of cloud modeling complexities on high SNR data of self-consistent models, and 2) identify an optimal set cloud parameters that fully describes the system but minimizes the number of free parameters. Using an identical modeling framework, \citet{nayak17} expanded on this work by exploring the effect of phase and radius uncertainty on the retrieval of atmospheric properties for cool giants with CH$_4$ dominated reflection spectra. Both of these studies concluded that optical spectra with a minimum SNR of 20 is required to retrieve accurate atmospheric properties like molecular abundances and clouds from reflected light for cool giants. 

\subsubsection{\citet{Feng_2018}}

\citet{Feng_2018} demonstrated the ability to ascertain atmospheric composition of earth-analog planets using higher resolution spectroscopy (R =70 or 140) as expected from mission concepts such as {\it HabEx} and {\it LUVOIR}. They considered only a single deck of H$_2$O cloud characterized by a cloud top pressure, cloud thickness and a single value of optical depth. The asymmetry parameter and single scattering albedo in their model is fixed to a constant value appropriate for H$_2$O clouds.  Abundances of O$_3$, O$_2$, H$_2$O and N$_2$ are retrieved, given the focus on Earth-like planet atmospheres. In addition to these parameters, they also retrieved a parameter for the reflective surface, and a parameter to describe the patchiness of clouds, {\it f$_c$}. Although their methodology pertains to terrestrial planets, the patchy cloud concept will likely be relevant for gas giants as well. 

In order to incorporate patchy cloud coverage, they create two models for each case: one with 100\% cloud coverage and a second that is cloud free. The albedo spectra of each of the runs are combined by weighting the first cloudy spectra with f$_c$ and the latter cloudless spectra with (1-{\it f$_c$}). They found that at relatively higher spectral resolutions of R$\sim$140 and SNR$\sim$20 the parameters of interest could be constrained. At R$\sim$70 and SNR$\sim$20 the presence of clouds and molecules could be detected and with low resolution (R $\sim$ 50) combined with SNR$\sim$20 one could achieve just weak detections of clouds and molecules. This retrieval model only accounts for water clouds. Terrestrial planet atmospheres might have photochemical hazes which would complicate inferring scattering properties with water only single deck cloud models. Lastly, \citet{feng2018earth} kept their single scattering albedo and asymmetry parameter fixed. This motivates additional work to determine how retrieving on these parameters for single or multiple cloud decks effects the retrieved solutions for cool giants. 

\subsubsection{\citet{Hu_19, Damiano2020exorel}}

A separate model, \texttt{ExoREL} was developed by \citet{Hu_19} for modeling reflected spectra of cool giants, and was implemented in a retrieval framework by \citet{Damiano2020exorel}. This model considered the effect of H$_2$O and NH$_3$ condensation on the vapor phase mixing ratio of these molecules.

\citet{Damiano2020exorel} conducted the retrieval analysis on three test planets, including {\it 47 Uma b}, a focus of this analysis as well. Similar to  \citet{lupu2016developing} and \citet{nayak17}, \citet{Damiano2020exorel} retrieved the cloud bottom pressure, cloud thickness, a well-mixed CH$_4$ mixing ratio and the gravity. But they assume only H$_2$O and NH$_3$ condensation and focused on retrieving the volume mixing ratio (VMR) of H$_2$O and NH$_3$ below the cloud bottom. They retrieve a condensation ratio which is then used to construct the depleted VMR of CH$_4$ and NH$_3$ above the cloud deck. \citet{Damiano2020exorel} retrieved on three test planet albedo spectra synthesized using \citet{Hu_19} forward model. 

\subsubsection{\citet{irwin2008nemesis, barstow2014clouds}}

\texttt{NEMESIS} is a well-vetted code for retreiving properties of Solar System planets \citep{irwin2008nemesis} and exoplanets \citet{barstow2014clouds}. Most recently, it was used to demonstrate the ability of retrieving cloud scattering properties in thermal emission \citep{Taylor20howdoesthermal}. With regards to reflected light, \citet{barstow2014clouds} used \texttt{NEMESIS} to determine the atmospheric parameters from hot Jupiter {\it HD 189733 b}, observed using HST/STIS by \citet{evans13} at very low spectral resolution. Instead of a Bayesian retrieval analysis, they perform a chi-square analysis on a grid of 980 spectra with a fixed set of cloud base pressures, particle sizes and optical depths at 0.25 microns -- and a fixed set of  chemistry  parameters for the volume mixing ratio of Na. Similar to previous analyses of cool giants where the cloud species is assumed, \citet{barstow2014clouds} assume that the clouds are composed of MgSiO$_3$ or MnS as these are the relevant condensation species for {\it HD 189733 b}. Then the scattering properties for the cloud particles were calculated using Mie theory with a double peaked Henyey–Greenstein formulation of the phase function. Other important atmospheric parameters such as the temperature-pressure profile and volume mixing ratios of CO, CH$_4$, H$_2$O, CO$_2$ were fixed at the best-fit values derived from previously obtained emission spectra. They found that the data was consistent with a large number of cloudy cases, as well as many cloudless cases. Therefore, there was lot of degeneracy in their cloud parameter space. Given the data quality and resolution of STIS, this is consistent with the findings of \citet{lupu2016developing,nayak2017atmospheric,feng2018earth} and \citet{Damiano2020exorel} who reported the need for a SNR of 20 for proper characterization of cloud properties.

\texttt{NEMESIS} has also been used for retrievals of scattering properties of Solar System planets from reflection spectra \citep[e.g.][]{irwin2015nemesis,irwin2016storm}. Unlike in the study of exoplanets, the cloud species, location and thickness for solar system planets are generally known quantities, and can be treated as fixed parameters. Moreover, the data quality used in these Solar System studies are generally far superior in quality than the focus of this and other previous studies discussed thus far. Therefore in studies such as \citet{irwin2015nemesis}, the wavelength-dependent imaginary part of the refractive indices and the parameters of the particle size distribution for the each condensing species can be directly retrieved from the data.

\subsection{This Analysis}
Previous work has highlighted the need for retrieval studies to be conducted on self-consistent models without the assumption of pure water clouds. Therefore, in this work, we build upon the approach of \citet{Lupu_2016} and others by retrieving the atmospheric properties of cool giant planets from spectra produced with a cloud model. We test four parameterizations to retrieve the cloud properties, each with increasing levels of complexity. Ultimately, we try to compare the accuracy and precision of the retrieved chemistry and clouds of the atmospheres to the original input. We focus on three radiative properties of the clouds -- optical depth per layer, asymmetry parameter and the single scattering albedo. By retrieving directly on the cloud radiative properties, we avoid any assumptions about the condensing species in the atmosphere. For our simulated data, we model the reflected spectra for three cool giants that are priority targets for {\it Nancy Grace Roman Space Telescope}. These priority cool giant targets have high contrast ratios compared to their host stars in reflected light due to the optimal combination of their size, separation from host star and effective temperature. 

 Table \ref{table:0} summarizes the spectral resolutions in the optical wavelength range expected from multiple future space-based direct imaging missions \citep{luvoir2019,habex20}. We consider both the lowest spectral resolution expected from {\it Nancy Grace Roman Space Telescope} and higher spectral resolutions expected from mission concepts like HabEx and LUVOIR while producing mock observations of the reflected spectra of these three priority target exoplanets. We retrieve on these mock observation spectra in order to test and compare various methodologies for parameterizing atmospheres when retrieving properties. This exercise helps to inform the complexity of atmospheric parameterization requisite for the next decade of reflected light studies. We also account for the uncertainty in the  gravity of the planets while performing the retrievals. In doing this, we aim to address the following:

\begin{enumerate}

%\item We demonstrate the retrieval capability of \texttt{PICASO} for cool giant planets

\item  Does the choice of cloud parameterization effect the retrieved cloud properties and molecular mixing ratios from the reflected spectra of cool giant planets?

\item Does the performance of our retrieval model change from one planet to another (i.e. different effective temperatures)? 

\item Does the constraint on the retrieved gravity depend on the cloud parameterization?

\item How does data quality (SNR \& R) limit ability to retrieve molecular abundances and cloud properties? 

\end{enumerate}
%\begin{savenotes}
\begin{table*}
\begin{center}

 \begin{tabular}{||c c c ||} 
 
 \hline
 Future Mission & Wavelength Range (microns) & Spectral Resolution \\ [0.5ex] 
 \hline\hline
 Roman Space Telescope CGI Spectroscopy &  0.675-0.785 & 47-75\footnote{\label{wfirst}\url{https://roman.ipac.caltech.edu/sims/Param_db.html}}  \\ 
 \hline
 Roman Space Telescope CGI Imaging & 0.5-0.8 \textsuperscript{\ref{wfirst}} & -  \\ 
 \hline
 LUVOIR-A \& B (ECLIPS) & 0.515-1.03 & 140 \footnote{\label{luvoir}\citet{luvoir2019}}  \\
 \hline
 HabEx Coronagraph & 0.45-1.00 & 140 \footnote{\label{habex}\citet{habex20}} \\%(logscale in Pressure) \\
 \hline
 HabEx Starshade Instrument & 0.45-0.975 & 140 \textsuperscript{\ref{habex}}\\%(logscale in Pressure) 
 \hline
\end{tabular}
\end{center}
\caption{Expected Spectral resolutions from future space-based direct imaging missions.}
\label{table:0}
\end{table*}
%\end{savenotes}

We describe the atmospheric modeling using \texttt{PICASO} and \texttt{Virga} in \S \ref{sec:Modeling}. We briefly explain the retrieval method in \S \ref{sec:nestedsampling}. We describe the results of our analysis in \S \ref{sec:results} and finally discuss our main results in \S \ref{sec:discussion} and summarize them in \S \ref{sec:conclusion}.
 
\section{Modeling Reflected Spectra}\label{sec:Modeling}
\begin{figure}
  \centering
  \includegraphics[width=0.45\textwidth]{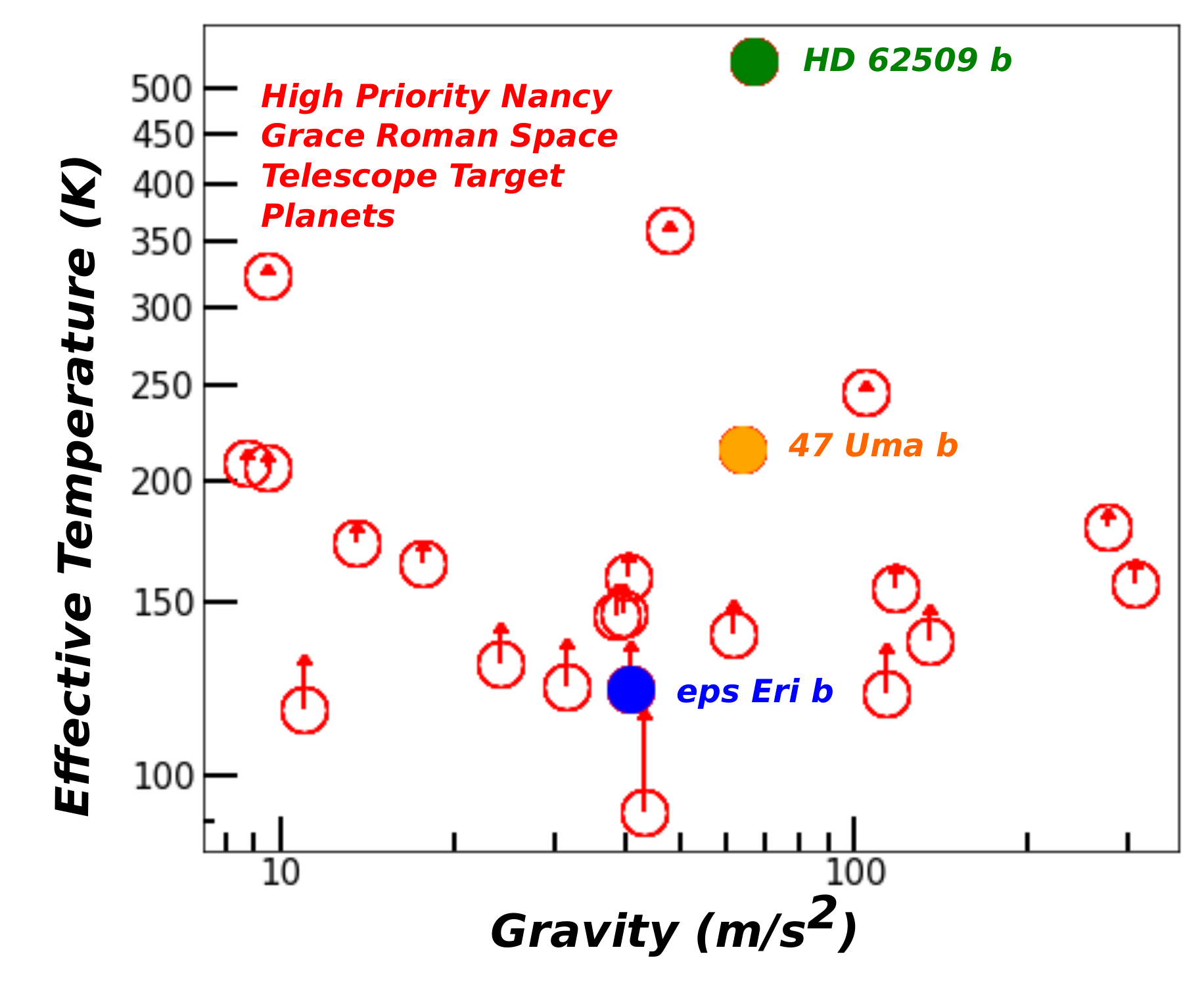}
  
  \caption{The effective temperature $T_{\rm eff}$ vs gravity $g$ parameter space for 23 cool giant targets for space-based direct imaging missions. The blue, orange and green points denote the target planets used in this work {\it eps Eri b}, {\it 47 Uma b} and {\it HD 62509 b}, respectively. The upward arrows point to the effective temperature if a 100 K internal temperature is considered. {\bf Main Point}- We choose three cool giant target planets that span a wide a range of effective temperatures (135~K - 533~K) in order to explore diversity in atmospheric cloud properties.}
\label{fig:fig1}
\end{figure}

We use the effective temperature vs. gravity parameter space to select a representative target population for the analysis. We calculate the equilibrium temperature ($T_{\rm eq}$) and gravity ($g$) of 23 direct imaging planet targets, most of which have radial velocity detections \citep{Butler_2006_catalog,howard16limits,fischer02second,hatzes06confirmation}. We use the planetary orbital and stellar parameters, to calculate the equilibrium temperature of these planets assuming a zero albedo. We also consider an additional internal temperature of 100 K, similar to that of Jupiter \citep{fortney2007planetary},  to get the effective temperature ($T_{\rm eff}$). %In reality this internal temperature will vary with planet mass and age. 
We use the $M \sin i$ and planet radius for calculating the gravity of these planets. We calculate the planet radius using the empirical mass-radius relationship for cool giants from \citet{Thorngren_2019empirical}. The $T_{\rm eff}$ vs $g$ parameter space for these planets are shown in Figure \ref{fig:fig1}. We use {\it eps Eri b}, {\it 47 Uma b} and {\it HD 62509 b} as our target planets (shown in in Figure \ref{fig:fig1}) to explore retrievals on reflected light across three different temperature regimes for directly imaged cool giants. We emphasize that the aim is not to produce highly self-consistent models of these planets. Instead, we aim to explore a range in temperature that enables a diversity in cloud formation, and chemistry scenarios. Additionally, our targets have similar gravity estimates. This allows us to isolate the effect that varying cloud and chemistry scenarios have on retrieving atmospheric parameters. In a future analysis we will explore the effect of completely unconstrained gravity. 

%%%%%%%%%%% THIS FIGURE HAS BEEN UPDATED WITH NEW RUNS

\subsection{Modeling the Planet Atmospheres}

We use \texttt{PICASO} \citep{batalha19}, which has heritage from \citet{mckay1989thermal, marley1999thermal,cahoy2010exoplanet}, to model the reflected light spectra of our target planets. \texttt{PICASO} is an open-source radiative transfer code capable of calculating transmission, reflected and/or thermal spectra of planets and brown dwarfs. \texttt{PICASO} requires the temperature-pressure ($T(P)$) profile, cloud structure and atmospheric chemistry as inputs for the radiative transfer calculation. Here, we discuss modeling each of these inputs for our reflected spectra simulation in \S\ref{sec:TP}-\S\ref{sec:cld} and discuss the basics of \texttt{PICASO} in \S\ref{sec:picaso}.

\subsubsection{The Temperature-Pressure Profile}\label{sec:TP}
\begin{figure}
  \centering
  \includegraphics[width=0.5\textwidth]{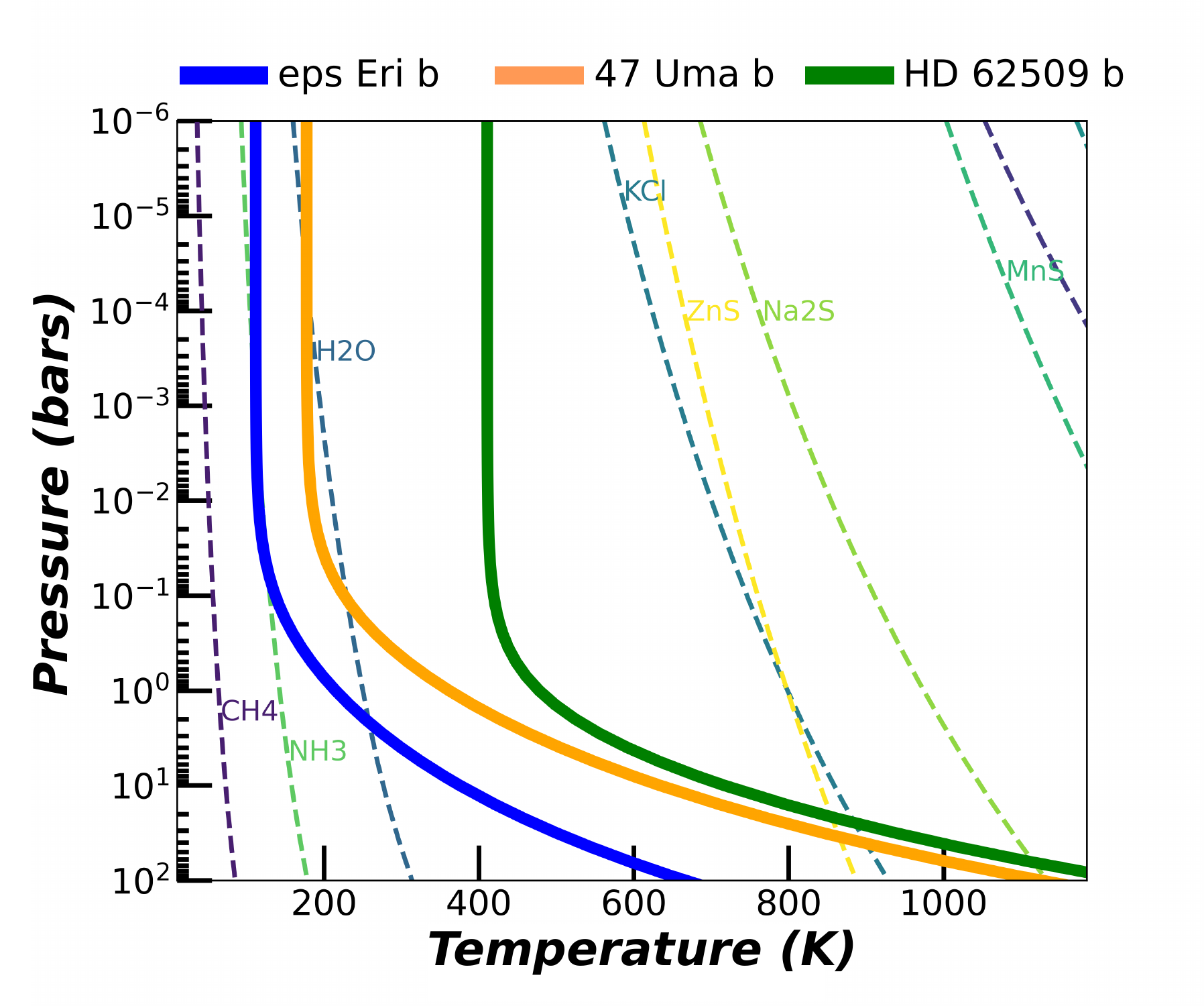}
  
  \caption{The temperature-pressure profile of {\it eps Eri b} (blue), {\it 47 Uma b} (orange) and {\it HD 62509 b} (green) generated using the empirical paramterization of \citet{ryan18} are shown. Condensation curves of the molecular condensates considered in the cloud calculation of \texttt{Virga} are shown by the colored dashed lines. {\bf Main Point}- Our target planets have different $T(P)$ profiles which causes condensation of different species at different pressures, resulting in varying cloud decks and optical properties.}
\label{fig:figtp}
\end{figure}

We divide the planet atmosphere in to 61 plane-parallel pressure layers where the pressure rises logarithmically from $10^{-6}$ to $10^{3}$  bars. We model the temperature-pressure profile of the planet atmospheres using the empirical parameterization described in \citet{ryan18}. This empirical $T(P)$ profile parameterization is dependent on T$_{\rm eff}$, $g$ and metallicity of the planet [M/H]. The best-fit value of the coefficients in the parameterization have been determined by fitting the empirical profile to a large number of $T(P)$ profiles for cool giants produced self-consistently using the methodology of \citet{fortney08}. The $T(P)$ profile parameterization is described by,
\begin{align}
    T^4(P) = {T_0}^4 + {T_{\rm deep}}^4 \left(\dfrac{P}{1000 \text{ bars}} \right)
\end{align}
where both $T_0$ and $T_{\rm deep}$ are T$_{\rm eff}$, $g$ and [M/H] dependent functions \cite[see][]{macdonald2018exploring}. We assume Jupiter metallicity of 3 $\times$ Solar metallicity \citep{wong04updated} for all the three planets. The parameterized $T(P)$ profile for the three planets are shown in Figure \ref{fig:figtp}. Although the self-consistent $T(P)$ profiles will show more structure than the parameterized profiles, as seen in \citet{macdonald2018exploring}, this will not affect the results of the analysis given the insensitivity of reflected light spectroscopy to temperature. 

%%%%%%%%%%% THIS FIGURE HAS BEEN UPDATED WITH NEW RUNS

\subsubsection{ Atmospheric Chemistry}\label{sec:chem}
\begin{figure*}
  \centering
  \includegraphics[width=1\textwidth]{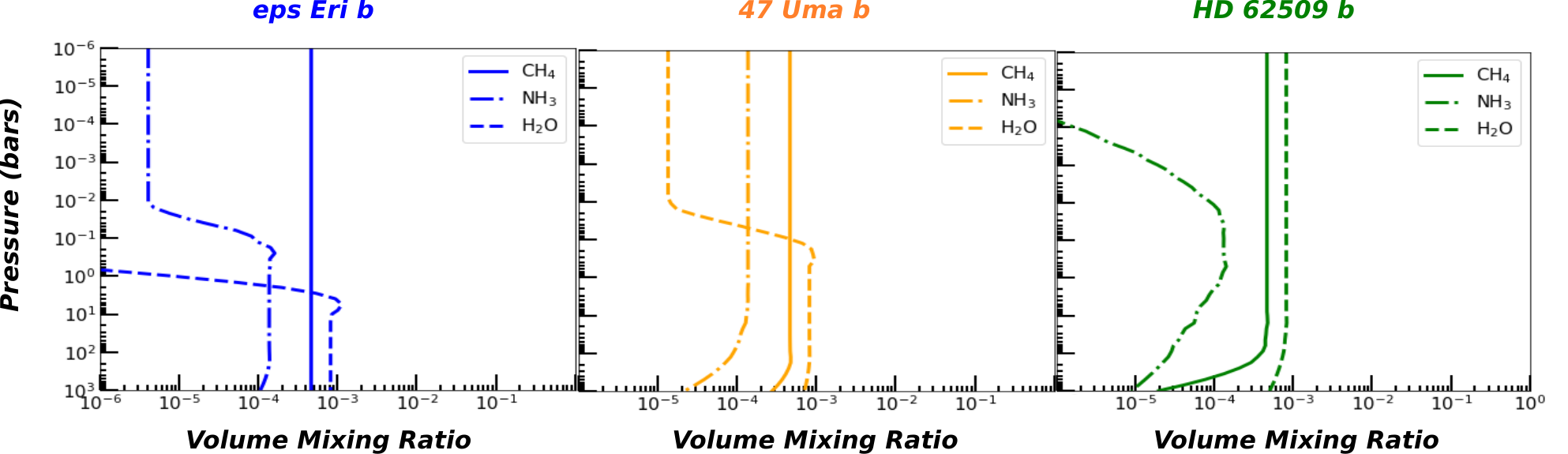}
  
  \caption{The volume mixing ratio as a function of pressure of CH$_4$, NH$_3$ and H$_2$O for the three planets.  {\bf Main Point}- The three planets have very different chemical structure due to different T(P) profiles. }
\label{fig:figchem}
\end{figure*}

Chemical equilibrium abundances are interpolated from those computed on a grid of $(P,T)$ points as calculated using a modified version of the NASA CEA Gibbs minimization code \cite[see][]{Gordon94}. The chemistry grid is available for download at \citet{marley_18} and described fully in Marley et al. (in prep.). For these cool giants the most important gaseous absorbers in the optical are methane, ammonia, and water. The grid accounts for depletion of each chemical species above the point of condensation. 
The volume mixing ratio profiles of these three species for each of our three target planets are shown in Figure \ref{fig:figchem}. 
%%%%%%%%%%% THIS FIGURE HAS BEEN UPDATED WITH NEW RUNS

\subsubsection{Clouds with \texttt{Virga}}\label{sec:cld}

We calculate the cloud profiles for each of the targets using \texttt{Virga}. \texttt{Virga} follows the \citet{marley00} treatment of condensation in atmospheres. For each condensate species and at each atmospheric layer, the vapor pressure in excess of the saturation vapor pressure is allowed to condense. The condensation curves of all the molecular species considered by our cloud model are shown in Figure \ref{fig:figtp} in dashed lines. The condensation curves follow those in \citet{morley2012neglected} and \citet{gao2020aerosol}, and are available online \footnote{\url{https://github.com/natashabatalha/virga/blob/master/virga/pvaps.py}}. 

\begin{figure*}
  \centering
  \includegraphics[width=1\textwidth]{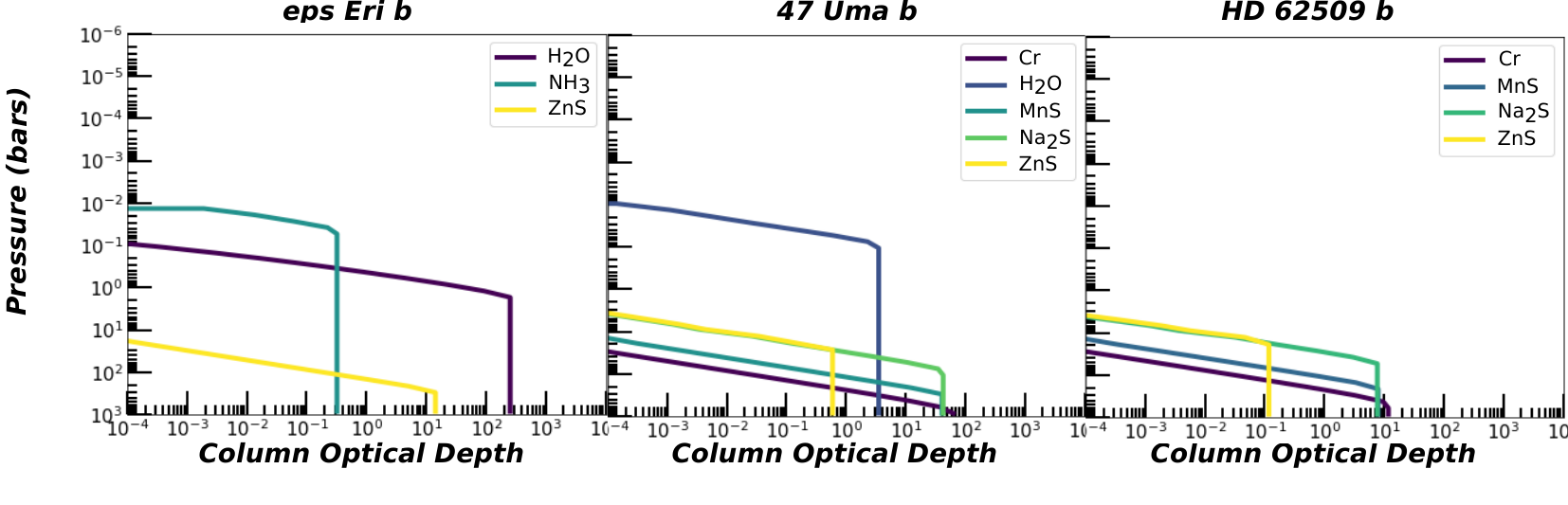}
  
  \caption{The column optical depth as a function of pressure for each condensed species for the three planets. {\it eps Eri b} has two major cloud decks of NH$_3$ and H$_2$O. The top deck for {\it 47 Uma b} is H$_2$O cloud while the bottom deck is composed primarily of Na$_2$S. {\it HD 62509 b} has a single cloud deck formed very deep in the atmosphere. Na$_2$S, ZnS and MnS are the high optical depth condensates among  other smaller optical depth condensate species for this planet. {\bf Main Point}- The three planets have very different cloud structure formed out of different condensate species. }
\label{fig:figcldopd}
\end{figure*}

The full cloud profiles from \texttt{Virga} are shown Figure \ref{fig:figcldopd}. The coolest planet case, {\it eps Eri b}, has H$_2$O and NH$_3$ condensation forming two separate cloud decks. The warmer case, {\it 47 Uma b}, lacks NH$_3$ clouds but is still dominated by H$_2$O clouds. The hottest case, {\it HD 62509 b}, is dominated by Na$_2$S clouds at depth (P$>1$~bar), and lacks condensation from H$_2$O or NH$_3$. As shown in Figure \ref{fig:figcldopd}, the three cases explored here probe three different cloud condensation regimes.
%%%%%%%%%%% THIS FIGURE HAS BEEN UPDATED WITH NEW RUNS

The vertical structure of the cloud is determined by the balance between the vertical turbulent mixing of condensates and vapor and the sedimentation of condensates described by the equation,

\begin{align}\label{eqn:am01}
    -K_z\dfrac{\partial q_t}{\partial z} - f_{\rm sed}w_*q_c=0
\end{align}

where $q_c$ is the condensate mole fraction, $q_v$ is the vapor mole fraction and $q_t=q_c+q_v$ \citep{marley00}. The first term represents the vertical turbulent mixing of the condensate and vapor, where $K_z$ is the vertical eddy diffusion coefficient. The second term represents the sedimentation caused by the condensates, where $w_*$ is the convective velocity scale. $f_{\rm sed}$ is a dimensionless ratio of the sedimentation velocity to $w_*$. The cloud structure is solved by the balance of the two competing processes for each of the condensing species. 

The sedimentation parameter $f_{\rm sed}$ and the vertical eddy diffusion coefficient $K_{z}$ are the two inputs which critically determine the cloud vertical extents and particle size distribution. Overall, low values of $f_{\rm sed}<1$ produce thick cloud layers with smaller particles and higher values, $f_{\rm sed}>1$, produce thinner cloud decks with larger particles. We set $f_{\rm sed} = 3$ motivated by the higher values that have been successful in modeling cool giant clouds for jupiter-like planets, in contrast to lower values of $\sim$0.1 which have been typically used for hot jupiters \citep{webber15}. 

The vertical eddy diffusion coefficient ($K_{z}$) can strongly effect the cloud properties. Overall, higher $K_{z}$ values lead to the formation of larger cloud particles. We calculate the vertical eddy diffusion coefficient $K_z$ using \citep{gierasch85energy},
\begin{align}
    K_z= \dfrac{H}{3}{\left(\dfrac{L}{H}\right)}^{\dfrac{4}{3}}{\left(\dfrac{RF}{\mu{\rho_a}c_p}\right)}^{\dfrac{1}{3}}
\end{align}

where $H$ is the atmospheric scale height, $L$ is turbulent mixing length, $R$ is the universal gas constant, $F$ is the thermal flux of the atmosphere (assumed to be ${\sigma}T_{\rm eff}^4$), $\mu$ is the atmospheric molecular weight, $\rho_a$ is the atmospheric density and $c_p$ is the atmospheric specific heat at constant pressure.

This formulation is based on the assumption that the vertical eddy diffusion coefficient for the vapor and the condensate of the cloud model is the same as derived for heat in free convection conditions \citep{gierasch85energy}. %A more complex treatment of $K_z$ is needed since 
This method assumes convection occurs all the way to the top of the atmosphere, which of course is not the case in reality. However, this has been used to baseline the cloud model for Jupiter in \citet{ackerman2001cloud} and, hence, is applicable for this study.

%Following the treatment of \citet{marley00}, we assume a lognormal distribution of cloud particle sizes at each pressure level where condensation has occurred. The lognormal distribution is described by two parameters namely r$_g$ -- the geometric mean radius of particles and $\sigma_g$ -- the geometric standard deviation (which we set to 2). These parameters are related to the f$_{sed}$ and $K_z$ through the formulation described in \citet{marley00}. With the vertical structure and particle size distribution of the condensation in the atmosphere, \texttt{Virga} employs Mie Scattering calculations using \texttt{PyMieScatt} \citep{sumlin18retrieving}.

In solving Equation \ref{eqn:am01}, we compute an effective particle radius per layer per species, and assume a lognormal distribution of particles with a geometric standard deviation of 2 about that radius. The Mie scattering calculations are then computed with \texttt{PyMieScatt} \citep{sumlin18retrieving} over this distribution. This allows \texttt{Virga} to produce the altitude- and wavelength-dependent optical depth per layer -- ${\tau}$(P,${\lambda})$, single scattering albedo -- $\omega$(P,$\lambda$)) and asymmetry parameter -- $g$(P,$\lambda$) of the clouds in the atmosphere. %The optical depth per layer captures the typical number of scatterings encountered by photons of a certain wavelength due to cloud particles at that atmospheric level. 
The single scattering albedo describes the wavelength-dependent reflectivity of the cloud particles. Higher $\omega$ leads to higher reflectivity. The asymmetry parameter captures the forward/back scattering probability of the scattering of light from the cloud particles. %The nature of the scattering phase functions can vary strongly with the asymmetry parameter of the cloud particles. 

\begin{figure*}
  \centering
  \includegraphics[width=1\textwidth]{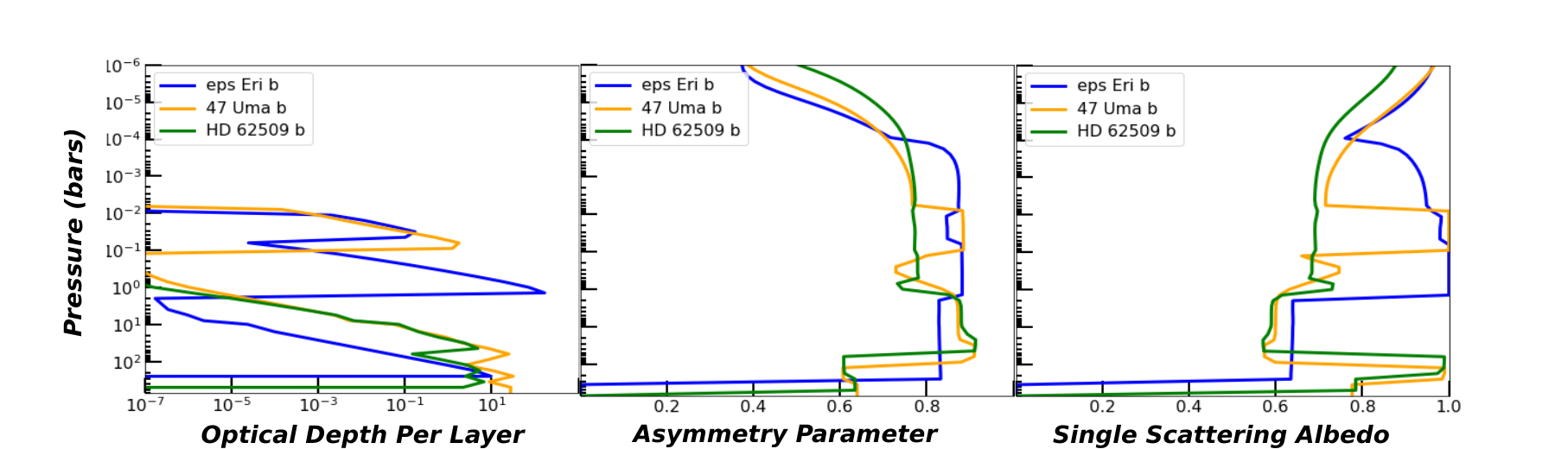}
  
  \caption{The cloud optical properties as outputs of the cloud modeling code \texttt{Virga} for the three planets. The optical depth per layer, asymmetry parameter and single scattering albedo are shown as function of pressure for the three planets from left to right, respectively. The cloud profiles shown here are averaged over the wavelength range of 0.3 to 1~$\mu$m. }
\label{fig:Virgacld}
\end{figure*}

The final cloud optical profiles of each planet are shown in Figure \ref{fig:Virgacld}. The optical properties can trace back to the exact cloud species. For example, the $\sim$1 bar cloud deck of eps Eri b corresponds to the highest region of single scattering and therefore, can be reasonably identified as a H$_2$O cloud. Ultimately, it is the information in these profiles that we aim to recover. 

%%%%%%%%%%% THIS FIGURE HAS BEEN UPDATED WITH NEW RUNS

\subsubsection{Reflected Spectra with \texttt{PICASO}}\label{sec:picaso}

With the $T(P)$ profile, chemical structure and the cloud optical properties as inputs, we adopt the one dimensional version of \texttt{PICASO} to calculate the reflected light spectra for the three planet cases \citep{batalha19}. \texttt{PICASO} uses the two-stream radiative transfer methodology of \citet{toon1989rapid}. We add  contributions from both molecular and collision-induced absorption (CIA) opacities. Although we only focus on retrieving H$_2$O, NH$_3$, and CH$_4$, \texttt{PICASO} includes the molecular opacity from H$_2$O \citep{barber06high,tennyson2018exomol}, CH$_4$ \citep{yurchenko_2014,yurchenko13vibrational}, NH$_3$ \citep{yurchenko11vibrationally}, CO \citep{li15rovibrational}, PH$_3$ \citep{sousa14exomol}, H$_2$S \citep{azzam16exomol}, CO$_2$ \citep{HUANG2014reliable}, Na \& K \citep{Ryabchikova2015}, and others not applicable to these temperatures (e.g. TiO, VO). Among the CIA \texttt{PICASO} includes opacity from H$_2$-H$_2$ \citep{collision-inducedmartin}, H$_2$-He, H$_2$-N$_2$, H$_2$-H, H$_2$-CH$_4$, H-electron bound-free, H-electron free-free and H$_2$-electron interactions. The resultant opacity calculations are available on Zenodo \citep{batalha_zenodo_opacities}.  

In order to accurately capture asymmetrical back scattering caused by Rayleigh scattering, we use the Two-Term Henyey-Greenstein (TTHG) phase function combined with Rayleigh phase function formalism (\texttt(TTHG$\_$Ray) in \texttt{PICASO}) for the direct scattering component. The One-Term Henyey-Greenstein (P$_{OTHG}$) phase function is defined as,
\begin{align}
P_{OTHG}(cos\Theta)=\dfrac{1}{2}\dfrac{1-g^2}{\sqrt{1+g^2 -2gcos{\Theta}}}     
\end{align}
The TTHG phase function capturing both forward, $g_f$, and back scattering, $g_b$, is then defined as
\begin{multline}
P_{TTHG}(cos\Theta)= fP_{OTHG}(cos{\Theta},g_f) \\ 
+(1-f)P_{OTHG}(cos{\Theta},g_b)\label{eqn:phase}
\end{multline}
where $g_f$= $\Bar{g}$, $g_b$= -$\Bar{g}$/2 and $f$, the fraction of forward to back scattering is $f= 1- g_{b}^2$. The first term corresponds to forward scattering phase function weighted by $f$ while the second term is the back scattering phase function weighted by $(1-f)$. Our adoption of $g_b$ is arbitrary. However, it has been previously adopted in studies of exoplanet reflected light \citep{cahoy2010exoplanet, Feng_2018} due to the lack of \textit{a priori} information. The $\Bar{g}$ is the cloud asymmetry parameter weighted by the cloud fractional opacity. It is calculated using the cloud asymmetry parameter, single scattering albedo, the cloud opacity, and Rayleigh opacity:
\begin{align}
\Bar{g}=\left(\dfrac{\omega\tau_{cld}}{\omega\tau_{cld}+\tau_{Ray}}\right)g
\end{align}
The TTHG and Rayleigh scattering phase functions ($P_{Ray}$) are then combined with a weighted addition to get the final phase function,  \texttt{TTHG$\_$Ray}. The weight factor for the TTHG and Rayleigh phase functions are $\tau_{cld}/\tau_{scat}$ and $\tau_{Ray}/\tau_{scat}$, respectively. P$_{TTHGRay}$ is then:
\begin{align}
P_{TTHGRay}(cos\Theta)= \dfrac{\tau_{cld}}{\tau_{scat}}P_{TTHG}(cos\Theta) + \dfrac{\tau_{Ray}}{\tau_{scat}}P_{Ray}(cos\Theta)
\end{align}

The multiple scattering phase function in \texttt{PICASO} is calculated by expanding the HG function to second order (\texttt{N=2}) and forcing the second order moment such that it reproduces Rayleigh scattering when it dominates the total scattering opacity \citep{batalha19}. We also include the effect of Raman scattering by using the \citet{POLLACK1986442} formalism. The \citet{POLLACK1986442} methodology for Raman scattering results in redshift of the photons, which dampens the overall reflectively toward the blue \citep{batalha19}. The \citet{POLLACK1986442} formulation does not model high resolution of solar emission features seen in reflected light spectra of gas giants \citep{antonija2016raman}. However, these features are far too high resolution ($R>>100$) for the next decade of direct imaging observations \citep{antonija2016raman}.  

\begin{figure*}
  \centering
  \includegraphics[width=1\textwidth]{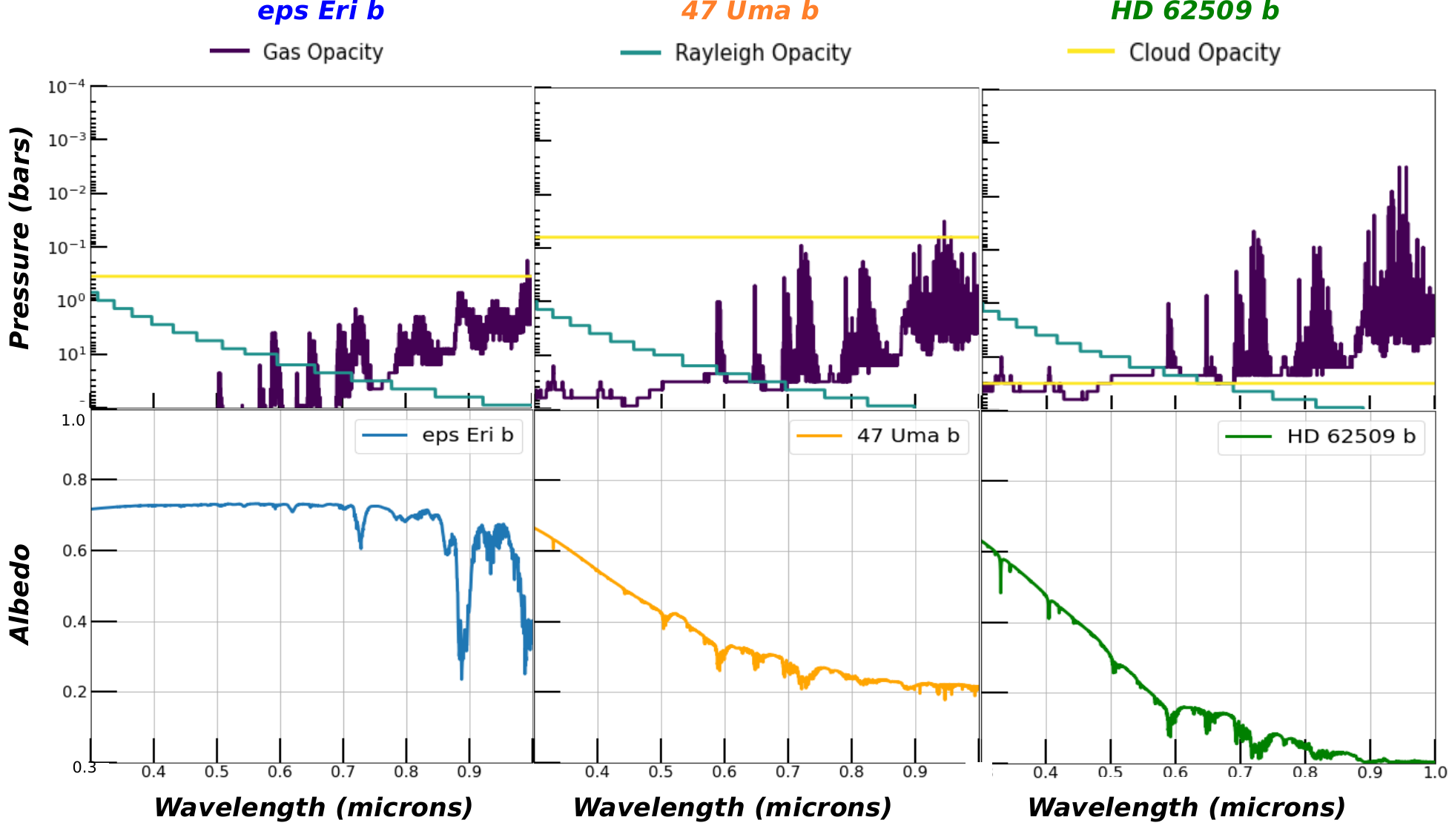}
  
  \caption{The photon attenuation depth maps in the top panel corresponding to an optical depth of 0.5 for our three fiducial cases: {\it eps Eri b} (left), {\it 47 Uma b} (middle) and {\it HD 62509 b} (right). The attenuation pressure levels are divided into Rayleigh, gas and cloud opacity. The lower panel shows the simulated albedo spectra for each planet. {\bf Main Point}- The three cases span three different scattering regimes. {\it eps Eri b}'s spectrum is dominated by bright water cloud reflection, {\it 47 Uma b}'s spectrum by cloud and molecular opacity, and {\it HD 62509 b}'s spectrum by molecular and Rayleigh scattering.}
\label{fig:figphot}
\end{figure*}

In order to understand the interplay between Rayleigh scattering, molecular, and cloud scattering, \texttt{PICASO} computes the ``photon attenuation'', which denotes the pressure level where the two-way optical depth from each component reaches $\tau=1$. Figure \ref{fig:figphot} shows the photon attenuation for the three planets cases. The ``flatness'' of the {\it eps Eri b} reflected spectra comes from the dominance of the water cloud optical properties, which are highly reflective and non-wavelength dependent at the respective particle radii from 0.3-1$\mu$m. On the other hand, the {\it HD 62509 b} spectrum is dominated by Rayleigh opacity short of 0.5~$\mu$m, by molecular absorption between 0.5-1$\mu$m. Because the cloud deck is much lower (in altitude) than the molecular opacity source, the molecular absorption dominates, causing the planet to have the lowest albedo among the three planet cases. Molecular, cloud and Rayleigh opacity contributes significantly to the {\it 47 Uma b} reflected light.

For the case of {\it eps Eri b} and  {\it 47 Uma b}, which both have contribution from the cloud, the different behaviour in the spectra is a result of different cloud optical properties (optical depth, single scattering, and asymmetry profiles). As seen from Figure \ref{fig:Virgacld} and Figure \ref{fig:figcldopd}, {\it eps Eri b} has two cloud decks (an NH$_3$ cloud deck above a very optically thick H$_2$O cloud deck), whereas {\it 47 Uma b} only has a H$_2$O cloud deck at a pressure similar to the higher NH$_3$ cloud deck in {\it eps Eri b}. The H$_2$O cloud deck in {\it eps Eri b} has a relatively high optical depth ($\sim$ 200) compared to the H$_2$O cloud deck in {\it 47 Uma b}. {\it 47 Uma b} also has a second cloud deck which appears much deeper in the atmosphere and is relatively optically thick compared to the deeper ZnS cloud deck appearing in {\it eps Eri b}. The optical depth differences, in addition to the differing asymmetry and the single scattering albedo of the {\it eps Eri b} H$_2$O cloud deck, leads to a flat cloud dominated albedo spectra for {\it eps Eri b} compared to {\it 47 Uma b}.

Finally, the simulated reflected light spectra for the three planets are shown in Figure \ref{fig:figphot} lower panel. They span: 1) a case dominated by bright water clouds, 2) a case dominated by clouds, molecular opacity and Rayleigh scattering, and lastly 3) a case dominated by Rayleigh and molecular opacity. These three spectra are exemplary cases to be used in the retrieval analysis as they test the parameterizations under three different scattering regimes.

%%%%%%%%%%% THIS FIGURE HAS BEEN UPDATED WITH NEW RUNS

\subsection{The Retrieval Setup}\label{sec:retrieval}

Similar to previous works discussed, a retrieval requires parameterizations to be made in order to best capture the behavior of the physical model, described in \S2.1. For the retrieval, we replace the chemistry, cloud, and $T(P)$ profile with parameterizations that can be used in a Bayesian framework. %We set up our model atmosphere into 61 plane-parallel pressure levels where the pressure rises logarithmically from 10$^{-6}$ bars to 10$^{2}$ bars. Similar to \citep{Lupu_2016}, 
We use an isothermal $T(P)$ profile for our forward model with the temperature fixed at the effective temperature of the planet. This is different from \citet{Lupu_2016} where the $T(P)$ profile for retrievals was kept fixed to the profile used for modeling the simulated data. Unlike thermal and transmission spectroscopy, reflected light is only sensitive to the temperature through its contribution to the line shapes of the molecular opacity, and the scale height of the atmosphere. The opacity is not strongly temperature dependent in the parameter space probed by our three targets \citep{karkoschka1994spec, karkoschka2011haze}. 

We initially assume that the atmosphere is well-mixed and start by retrieving a single value for the mixing ratios of three molecules -- CH$_4$, H$_2$O and NH$_3$ -- that dominate the opacity sources for these cooler class of planets \citep{madhusudhan16,burrows97}. The rest of the atmospheric composition, other than CH$_4$, NH$_3$ \& H$_2$O, is assumed to be composed of H$_2$ and He. The H$_2$/He fraction is taken to be f=H$_2$/He=$0.837/0.163$ \citep{lodders19solar}. Therefore the He and H$_2$ mixing ratio is given by,
\begin{align}
    He &= \dfrac{1- CH_4- H_2O-NH_3}{f+1},\\        
    H_2  &= f*He
\end{align}

where the species' name represent its volume mixing ratios (v/v). Our initial assumption of well-mixed profiles in the retrieval is different from that of \citet{Damiano2020exorel}, who use two free parameters for H$_2$O and NH$_3$ each to describe their depleted mixing ratios. %The CH$_4$ mixing ratio is assumed to be well mixed in \citet{Damiano2020exorel} similar to this study. 
Our chemical profiles incorporate depletion caused by condensation as is clearly evident in Figure \ref{fig:figchem}. {\it eps Eri b} shows a depletion of NH$_3$ and H$_2$O due to condensation of both the species whereas {\it 47 Uma b} shows a depletion only in the H$_2$O mixing ratio as NH$_3$ condensation is absent in {\it 47 Uma b}. Therefore, our initial assumption of well-mixed atmospheres purposely tests whether or not additional complexity is necessary. Later, we relax this assumption and discuss the results of a retrieved depleted profile in \S\ref{sec:depletedwater}.

\begin{figure}
  \centering
  \includegraphics[width=0.5\textwidth]{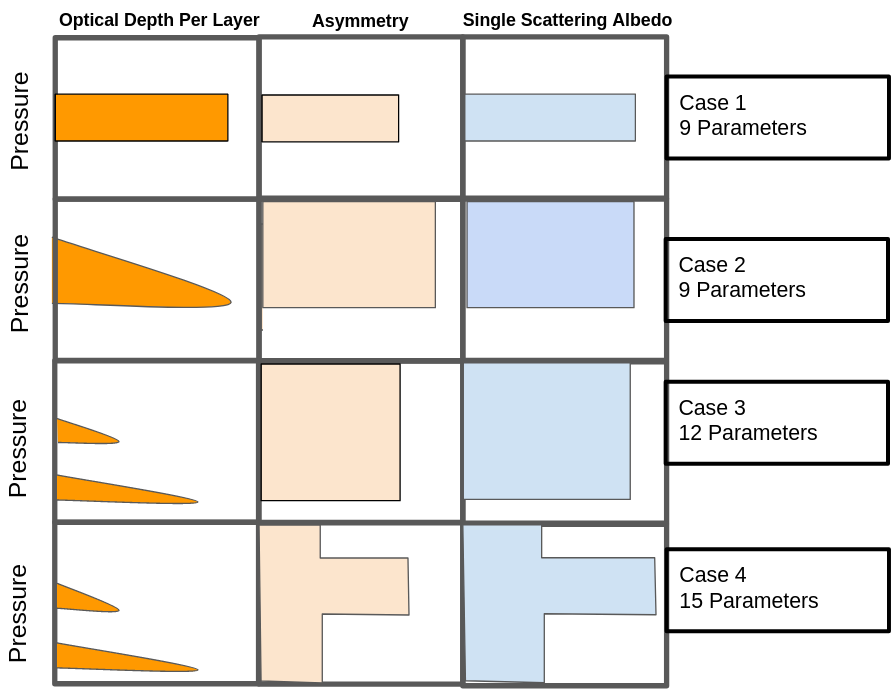}
  
  \caption{Schematic diagram of the four cloud structure and optics parameterizations used in our forward model for the target planets. The left column shows the optical depth per layer -- $\tau$, the middle column is for the asymmetry parameter -- {\it g} per layer and the right column shows the single scattering albedo -- $\omega$ per layer. The top row shows the parameterizations for Case 1 -- the box cloud model, the second row is for Case 2 -- the single cloud profile model, the third row shows Case 3 -- the double cloud profile model and the last row shows Case 4 -- the double cloud profile model with two valued asymmetry and single scattering albedo model. Details of the model parameters for each case can be found in \S \ref{sec:boxcld} through \S  \ref{sec:case4}. %{\bf Main Point}- Four different parameterizations for the cloud model are used to test their dependence on retrieved atmospheric properties.
  }
\label{fig:fig3}
\end{figure}

We parameterize ${\tau}$(P), $\omega$(P) and $g$(P) in four ways with different levels of complexities. We neglect any wavelength dependence in all the three cloud optical properties for our retrieval model. The validity of this assumption, given that \texttt{Virga} calculates wavelength dependent cloud optical properties, is addressed in  \S\ref{sec:discussion}. The schematic diagram for the cloud parameterizations are shown in Figure \ref{fig:fig3}. In what follows, we describe each of the four parameterizations (9-15 free parameters total), and the associated priors used in the retrieval analysis.

\subsubsection{The Box Cloud Model (Case 1)}\label{sec:boxcld}

The optical depth profile ($\tau(P)$) of this cloud parameterization is similar to the cloud deck used for retrieval by \citet{Feng_2018}. Unlike \citet{Feng_2018}, the asymmetry parameter and the single scattering albedo for this parameterization are also free parameters. We model the cloud structure in this case using 5 parameters. The other four parameters are the CH$_4$, NH$_3$ and H$_2$O mixing ratios and the gravity of the planet. The cloud structure for this case is parameterized according to the following equations,
\begin{align}
    \tau (P,\lambda) =
    \begin{cases}
        0 & \text{if $P > P_0$ or $P < P_0 - dP$ ,}\\
        \tau_{max} & \text{otherwise}
    \end{cases}
\end{align}

\begin{align}
    \omega (P,\lambda) =
    \begin{cases}
        0 & \text{if $P > P_0$ or $P < P_0 - dP$ ,}\\
        \omega & \text{otherwise}
    \end{cases}
\end{align}

\begin{align}
    g (P,\lambda) =
    \begin{cases}
        0 & \text{if $P > P_0$ or $P < P_0 - dP$ ,}\\
        g & \text{otherwise}
    \end{cases}
\end{align}
%\end{equation} 

\smallskip
where P$_0$, $\tau_{max}$, $\omega$, g and dP are the 5 parameters of the model. The nine parameters of this model and the priors used in retrievals are summarised in Table \ref{table:1}.
\begin{table*}
\begin{center}

 \begin{tabular}{||c c c c||} 
 
 \hline
 Parameter & Description & Range & Iteration type \\ [0.5ex] 
 \hline\hline
 g & Asymmetry Parameter & 0-1 & Linear \\ 
 \hline
 $\omega$ & Single Scattering Albedo & 0-1 & Linear \\
 \hline
 $\tau$ & Optical Depth Per Layer & 0.1-30 & Linear \\
 \hline
 P$_0$ & Cloud Base Pressure Level & 10$^{-6}$-10$^2$ bars  & Logscale \\%(logscale in Pressure) \\
 \hline
 dP & Cloud Deck Thickness & 0 to $P_0$-10$^{-6}$ bars & Logscale \\%(logscale in Pressure) \\
 \hline
 CH$_4$ & CH$_4$ Mixing Ratio & -6 to 0 & Logscale \\
 \hline
 NH$_3$ & NH$_3$ Mixing Ratio & -6 to 0 & Logscale \\
 \hline
 H$_2$O & H$_2$O Mixing Ratio & -6 to 0 & Logscale \\
 \hline
 g & Gravity  & 25-65 m/s$^2$ & Linear \\ [1ex] 
 \hline
\end{tabular}
\end{center}
\caption{Parameters for the box cloud model.}
\label{table:1}
\end{table*}

\subsubsection{Single Cloud Profile Model (Case 2)}\label{sec:profcld}

For this case, we model the clouds with an altitude dependent optical depth profile. The asymmetry parameter and the single scattering albedo are forced to be zero beneath the base of the cloud deck and can take a value between zero and one above the base of the cloud deck till the top of the atmosphere. The cloud structure here is modeled as,

\begin{align}
    \tau (P,\lambda) =
    \begin{cases}
        0 & \text{if $P > P_0$ ,}\\
        \tau_{max} {\rm e}^{-a(lnP-lnP_0)^2} & \text{otherwise}
    \end{cases}
\end{align}

\begin{align}
    \omega (P,\lambda) =
    \begin{cases}
        0 & \text{if $P > P_0$ ,}\\
        \omega & \text{otherwise}
    \end{cases}
\end{align}

\begin{align}
    g (P,\lambda) =
    \begin{cases}
        0 & \text{if $P > P_0$ ,}\\
        g & \text{otherwise}
    \end{cases}
\end{align}

where P$_0$, $\tau_{max}$, a, $\omega$ and g are free parameters. Hence, our forward model involves 9 free parameters consisting of 5 cloud parameters and the mixing ratios and gravity, similar to Case 1. The number of parameters are same as Case 1 and they are described in Table \ref{table:2}.

\begin{table*}
\begin{center}

 \begin{tabular}{||c c c c||} 
 
 \hline
 Parameter & Description & Range & Iteration type \\ [0.5ex] 
 \hline\hline
 g & Asymmetry Parameter & 0-1 & Linear \\ 
 \hline
 $\omega$ & Single Scattering Albedo & 0-1 & Linear \\
 \hline
 $\tau$ & Optical Depth Per Layer & 0.1-30 & Linear \\
 \hline
 P$_0$ & Cloud Base Pressure & 10$^{-6}$-10$^2$ bars  & Logscale  \\
 \hline
 a & Cloud Deck Scale Height & 10$^{-4}$ to 2 & Logscale \\
 \hline
 CH$_4$ & CH$_4$ Mixing Ratio & -6 to 0 & Logscale \\
 \hline
 NH$_3$ & NH$_3$ Mixing Ratio & -6 to 0 & Logscale \\
 \hline
 H$_2$O & H$_2$O Mixing Ratio & -6 to 0 & Logscale \\
 \hline
 g & Gravity  & 25-65 m/s$^2$ & Linear \\ [1ex] 
 \hline
\end{tabular}
\end{center}
\caption{Parameters for the single cloud profile model.}
\label{table:2}
\end{table*}

\subsubsection{Double Cloud Profile Model (Case 3)}\label{sec:case3}

This model is similar to Case 2 except a second cloud deck is allowed to form here. Additionally, the asymmetry and single scattering profiles are similar to Case 2 where they take a value between zero and one above the base of the deepest cloud deck.  The twelve free parameters are described in Table \ref{table:3}. The following equations describes the parameterizations for this model,

\begin{align}
    \tau (P,\lambda) =
    \begin{cases}
        0 & \text{if $P > P_1$ ,}\\
        \tau_{max1} {\rm e}^{-a_1(lnP-lnP_1)^2} & \text{if $P < P_1$ ,} \\
        \tau_{max2} {\rm e}^{-a_2(lnP-lnP_2)^2} & \text{if $P < P_2$ ,} \\
    \end{cases}
\end{align}

\begin{align}
    \omega (P,\lambda) =
    \begin{cases}
        0 & \text{if $P > P_1$ ,}\\
        \omega & \text{otherwise}
    \end{cases}
\end{align}

\begin{align}
    g (P,\lambda) =
    \begin{cases}
        0 & \text{if $P > P_1$ ,}\\
        g & \text{otherwise}
    \end{cases}
\end{align}

\begin{table*}
\begin{center}

 \begin{tabular}{||c c c c||} 
 
 \hline
 Parameter & Description & Range & Iteration type \\ [0.5ex] 
 \hline\hline
 g & Asymmetry Parameter & 0-1 & Linear \\ 
 \hline
 $\omega$ & Single Scattering Albedo & 0-1 & Linear \\
 \hline
 $\tau_1$ & Optical Depth Per Layer of Lower Cloud Deck & 0.1-30 & Linear \\
 \hline
 $\tau_2$ & Optical Depth Per Layer of Upper Cloud Deck & 0.1-30 & Linear \\
 \hline
 P$_1$ & Cloud Base Pressure of Lower Cloud Deck & 10$^{-6}$-10$^2$ bars  & Logscale  \\
  \hline
 P$_2$ & Cloud Base Pressure of Upper Cloud Deck & P$_1$-10$^{-6}$ bars  & Logscale\\
 \hline
 a$_1$ & Cloud Deck Scale Height of Lower Cloud Deck & 10$^{-4}$ to 2 & Logscale \\
 \hline
 a$_2$ & Cloud Deck Scale Height of Upper Cloud Deck & 10$^{-4}$ to 2 & Logscale \\
 \hline
 CH$_4$ & CH$_4$ Mixing Ratio & -6 to 0 & Logscale \\
 \hline
 NH$_3$ & NH$_3$ Mixing Ratio & -6 to 0 & Logscale \\
 \hline
 H$_2$O & H$_2$O Mixing Ratio & -6 to 0 & Logscale \\
 \hline
 g & Gravity  & 25-65 m/s$^2$ & Linear \\ [1ex] 
 \hline
\end{tabular}
\end{center}
\caption{Parameters for the double deck cloud profile model.}
\label{table:3}
\end{table*}

\subsubsection{Double Cloud Profile Model with Two Valued g0 and w0 (Case 4)}\label{sec:case4}
This fifteen parameter model has the same optical depth per layer parameterization as Case 3. The major difference for this case is that the asymmetry and single scattering are each allowed to have two-values. The asymmetry and single scattering parameterizations are,

\begin{align}
    \omega (P,\lambda) =
    \begin{cases}
        \omega_2 & \text{if $P > P_2$ and $P < P_2-dP$ ,}\\
        \omega_1 & \text{otherwise}
    \end{cases}
\end{align}

\begin{align}
    g (P,\lambda) =
    \begin{cases}
        g_2 & \text{if $P > P_2$ and $P < P_2-dP$ ,}\\
        g_1 & \text{otherwise}
    \end{cases}
\end{align}

Table \ref{table:4} describes the 15 parameters for Case 4.

\begin{table*}
\begin{center}

 \begin{tabular}{||c c c c||} 
 
 \hline
 Parameter & Description & Range & Iteration type \\ [0.5ex] 
 \hline\hline
 $g_1$ & Asymmetry Parameter & 0-1 & Linear \\ 
 \hline
 $g_2$ & Asymmetry Parameter & 0-1 & Linear \\ 
 \hline
 $\omega_1$ & Single Scattering Albedo & 0-1 & Linear \\
 \hline
 $\omega_2$ & Single Scattering Albedo & 0-1 & Linear \\
 \hline
 $\tau_1$ & Optical Depth Per Layer of Lower Cloud Deck & 0.1-30 & Linear \\
 \hline
 $\tau_2$ & Optical Depth Per Layer of Upper Cloud Deck & 0.1-30 & Linear \\
 \hline
 P$_1$ & Cloud Base Pressure Level of Lower Cloud Deck & 10$^{-6}$-10$^2$ bars  & Logscale  \\
  \hline
 P$_2$ & Cloud Base Pressure Level of Upper Cloud Deck & P$_1$-10$^{-6}$ bars  & Logscale \\
 \hline
 a$_1$ & Cloud Deck Scale Height of Lower Cloud Deck & 10$^{-4}$ to 2 & Logscale \\
 \hline
 a$_2$ & Cloud Deck Scale Height of Upper Cloud Deck & 10$^{-4}$ to 2 & Logscale \\
 \hline
 dP & Thickness of g0/w0 Deck & 0 to P$_2$-10$^{-6}$ bars & Logscale \\
 \hline
 CH$_4$ & CH$_4$ Mixing Ratio & -6 to 0 & Logscale \\
 \hline
 NH$_3$ & NH$_3$ Mixing Ratio & -6 to 0 & Logscale \\
 \hline
 H$_2$O & H$_2$O Mixing Ratio & -6 to 0 & Logscale \\
 \hline
 g & Gravity  & 25-65 m/s$^2$ & Linear \\ [1ex] 
 \hline
\end{tabular}
\end{center}
\caption{Parameters for the double deck cloud profile model with two valued $g$ and $w$.}
\label{table:4}
\end{table*}

\section{Dynamic Nested Sampling}\label{sec:nestedsampling}
We use the Dynamic Nested Sampling package \texttt{Dynesty}{\footnote{\url{https://dynesty.readthedocs.io/en/latest/}}} \citep{speagle20} for our retrievals. We choose Dynamic Nested Sampling to explore non-gaussian posteriors of the parameters. The Nested Sampling \citep{skilling2006} method can efficiently and accurately determine both the evidence and the posteriors of the problem simultaneously, unlike traditional MCMC that prioritizes the estimation of the posterior. Briefly, the basic process includes: 1) drawing a large number ($\sim 50\times$ number of free parameters) of live points from the provided priors of the parameter space, and then 2) iteratively replacing the live point with the least likelihood with a new live point having a greater likelihood than the replaced point. At each iteration, the replaced points become ``dead points''. The evidence can then be estimated with a set of $N$ dead points by summing over the product of their likelihoods and prior volumes \citep{skilling2006}. This process continues until a user-defined stopping criteria is met. 

We wrap the retrieval model described in \S\ref{sec:retrieval} in the \texttt{Dynesty} module and retrieve on the simulated observational spectra. In each retrieval, we assign 50 live points per free parameter, as recommended \citep{speagle20}. We use the the multi- ellipsoidal decomposition method because of its ability to efficiently capture complex, multi-modal posteriors \citep{speagle20}. We use ($\Delta ln(Z_i)$) defined as,
\begin{align}
  \Delta ln(Z_i) = ln(Z_i+\Delta Z_i)-ln(Z_i)  
\end{align}
as our stopping criteria. Here, $Z_i$ and $\Delta Z_i$ are the current and remaining evidence estimate, respectively. The remaining evidence can be approximated by the product of the highest likelihood among remaining live-points and the prior volume of the last dead point. The retrieval is stopped when this $\Delta ln(Z_i)$ is smaller than $N/1000$ where $N$ is the number of live points. This criteria has been optimized for evidence and posterior estimation by \citet{speagle20}. We do not use any limitation on the maximum number of iterations for the retrievals. 

We use the evidences estimated from the nested sampling for calculating the Bayes factor. The Bayes factor allows us to quantitatively compare each of our models. This directly can inform us whether or not one model is favored, compared to another. The Bayes factor of model M$_0$ over model M$_1$ for a dataset $D$ is,
\begin{align}
    B_{01} = \dfrac{p(D|M_0)}{p(D|M_1)}
\end{align}

where $p(D|M_0)$ and $p(D|M_1)$ are the evidence of model 0 and model 1 with the dataset $D$, respectively. Pairs of models with $\ln(B)$ less than 2.1 are said to indicate that model 0 is weakly favored over model 1 at best with a confidence of less than 2$\sigma$ \citep{Trotta_2008}. If $\ln(B)$ is greater than 5, model 1 can be strongly ruled out over model 0 with a confidence of 3.6-5$\sigma$ \citep{Trotta_2008}.
We present the retrieval results obtained using \texttt{Dynesty} in the following sections.  

\section{Results}\label{sec:results}

We first bin the model spectra for {\it eps Eri b}, {\it 47 Uma b}, and {\it HD 62509 b} to a constant resolution (R) of 40. We also fix the signal-to-noise ratio (SNR) of the spectra to 20 at a wavelength of 0.35 microns. This initial choice of SNR is motivated by earlier studies \citep{Lupu_2016,nayak17,Hu_19,Feng_2018} establishing this to be the minimum SNR required for accurate retrievals of atmospheric properties. The values for the R and SNR are chosen to mimic the likely best possible data quality from near future space-based direct imaging and spectroscopy missions. We also explore the effect of degrading SNR to 5. We retrieve on the albedo spectra for each planet using our four parameterizations described in \S \ref{sec:retrieval}. Here we present the results of our analysis on a planet-by-planet basis.

\subsection{47 Uma b}

\subsubsection{Comparison of Retrieval parameterizations}\label{sec:47umab}

\begin{figure}
  \centering
  \includegraphics[width=0.5\textwidth]{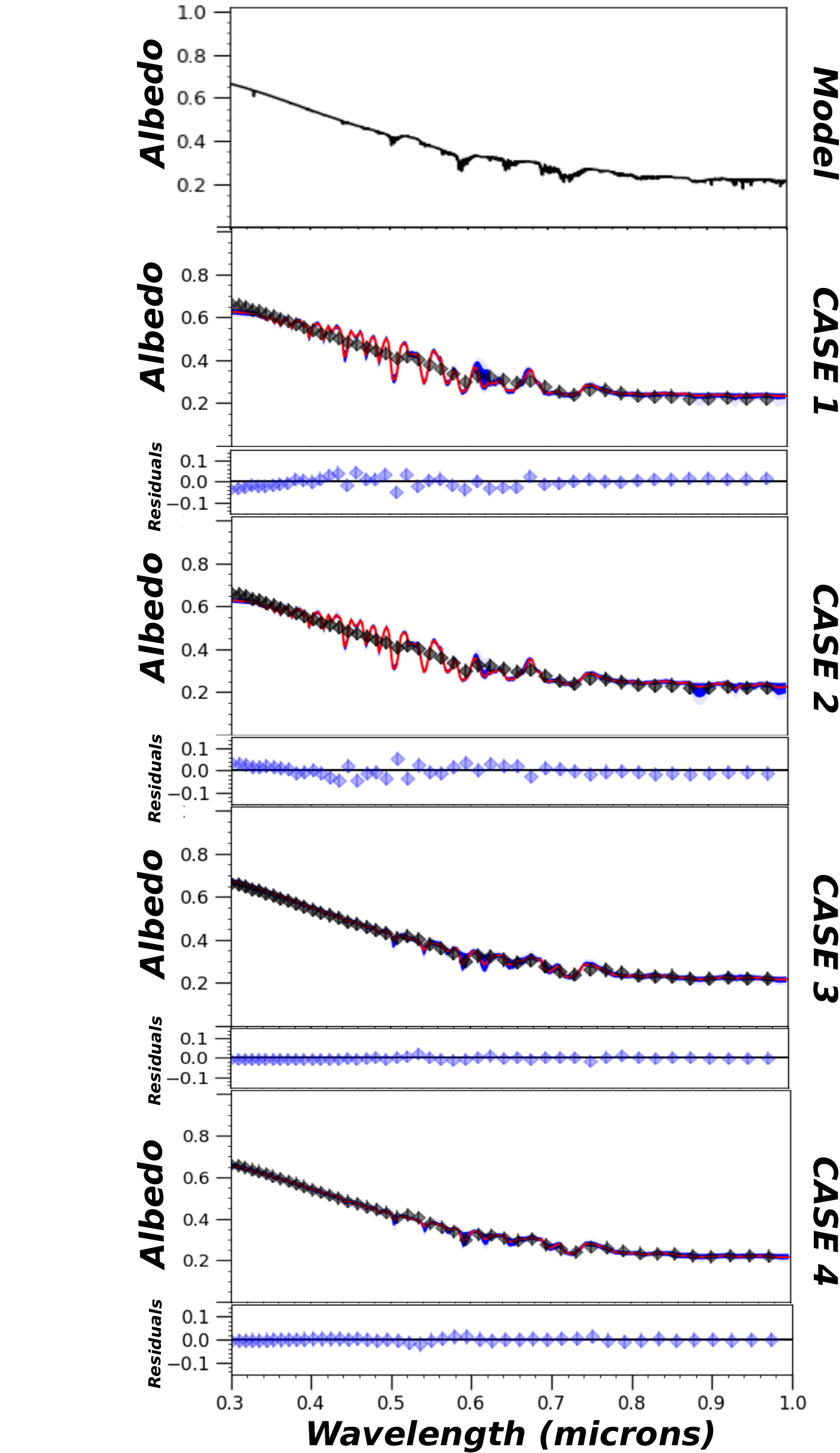}
  
  \caption{The top panel shows the model albedo spectrum for {\it 47 Uma b}. The second,third, fourth and fifth panels show the comparison of retrieved spectra by each retrieval model from Case 1 to Case 4 with the simulated observed spectra of {\it 47 Uma b} shown in black diamonds. The red line shows the median albedo spectra of the retrieved models while the dark and light blue shaded regions represent the 1$\sigma$ and 2$\sigma$, respectively. {\bf Main Point}- Case 3 and 4 provide much better fits to the spectra than Cases 1 and 2.}
\label{fig:fig4}
\end{figure}

For the first case, we retrieve atmospheric parameters using all four cases described in \S \ref{sec:retrieval} on {\it 47 Uma b}. We estimate the effective temperature of {\it 47 Uma b} to be $\sim$ 217 K. Figure \ref{fig:fig4} shows the median retrieved solution along with 1$\sigma$ and 2$\sigma$ confidence intervals. The residuals of the retrieved median spectra from the simulated data are also shown for each case in Figure \ref{fig:fig4}. 

\begin{figure}
  \centering
  \includegraphics[width=0.5\textwidth]{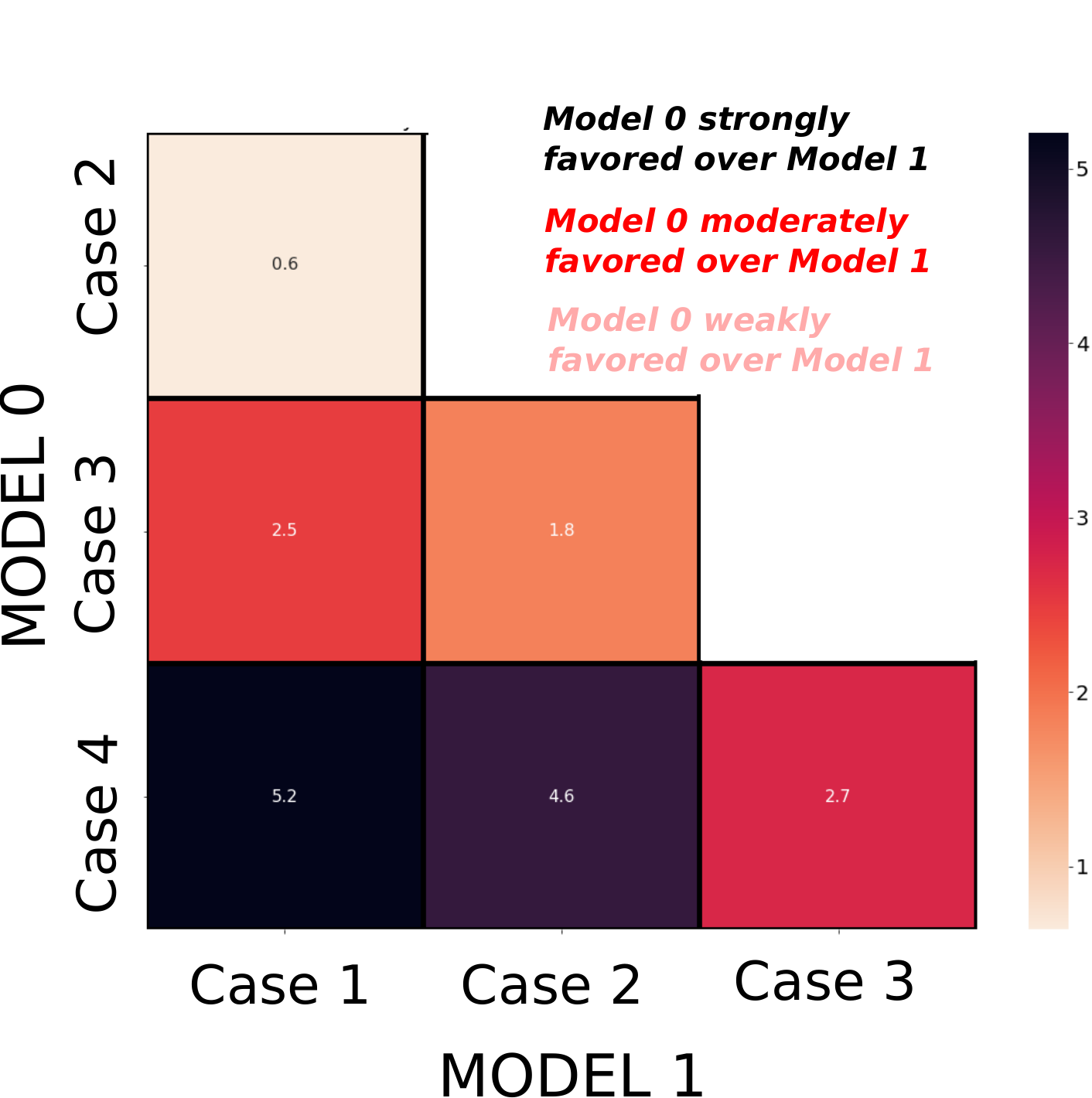}
  
  \caption{Heat map of the natural logarithm of the Bayes Factor of the ``Model 0'' on the y-axis over ``Model 1'' on the x-axis. A higher Bayes factor, as is seen for Case 3 and 4 compared to Case 1 and 2, signifies that Case 3 and 4 can be favored over Case 1 and 2. Between Case 1 and 2, Case 2 very weakly rejects Case 1. Between Case 3 and 4, Case 4 moderately rejects Case 3. {\bf Main Point}- Case 3 and 4 outperform Case 1 and 2 in retrievals. Case 4 (highest complexity) is moderately favored over Case 3.}
\label{fig:bayes}
\end{figure}

Case 1 and Case 2 show greater residuals on the blue side of the spectra compared to Case 3 and 4. The large residuals toward the blue for the  Case 1 and 2 retrievals foreshadows a potential overestimation of the molecular abundances. However, comparing the performances of Case 3 and Case 4 just with the residuals and the median spectra is not quantitatively informative. Hence, we use the evidence estimates from the nested sampling calculations in order to determine how strongly one model could be ruled out or compared to another model. Specifically, we calculate the Bayes factor described in \S \ref{sec:nestedsampling} for each pair of models. The heat map is shown in Figure \ref{fig:bayes}. As suggested by Figure \ref{fig:fig4}, we see that both Case 3 and Case 4 are favored over Case 1 and Case 2. Between the two 9 parameter models, Case 2 is favored very weakly over Case 1. The 15 parameter model Case 4 is moderately favored over the 12 parameter model Case 3. Moving forward we evaluate each of the retrieval models by comparing their retrievals of various atmospheric properties with the input properties used to simulate the mock spectra. 

%think this was a bit out of place.. 
%Comparing the retrieved photon attenuation maps for all the cases with the `true' photon attenuation map for {\it 47 Uma b} shown in Figure \ref{fig:figphot} shows that the retrieved cloud contributions are placed higher in altitude in Case 1 and Case 2 than the cloud contribution in the 'true' photon attenuation map. The molecular opacity contributions are also overestimated in the retrieved model for Case 1 and 2 which causes large number of additional absorption features towards the blue side. The retrieved photon attenuation map for Case 3 and Case 4 places the cloud component higher than the `true' photon attenuation map although the Rayleigh and gaseous components  match between the 'true' and retrieved models. 

%%%%%%%%%%% THIS FIGURE HAS BEEN UPDATED WITH NEW RUNS

\begin{figure}
  \centering
  \includegraphics[width=0.5\textwidth]{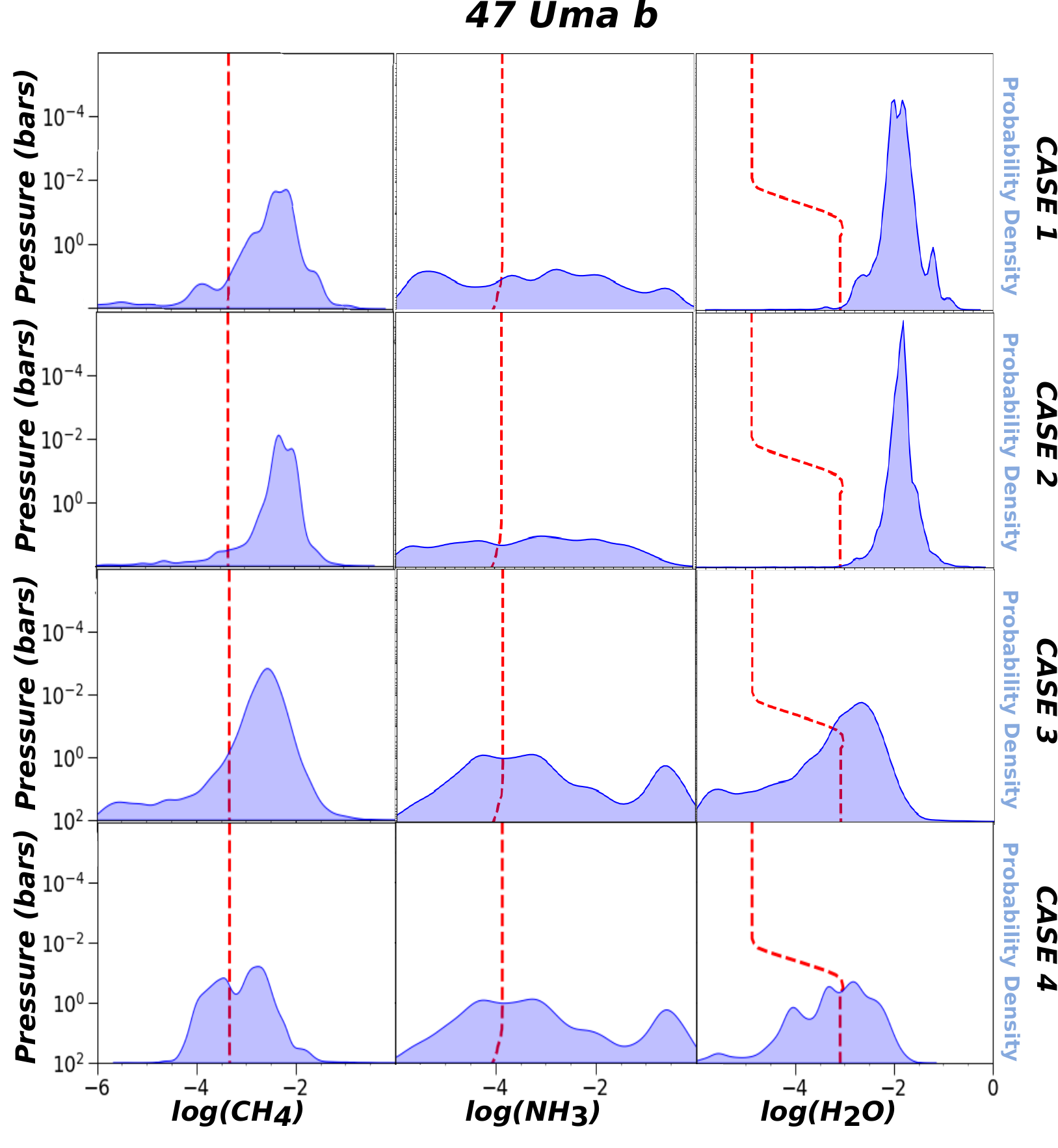}

  \caption{The retrieved molecular mixing ratios of CH$_4$, NH$_3$ and H$_2$O with the input mixing ratio profile used to generate the `observed' albedo spectra for {\it 47 Uma b}. The four rows depict the retrievals for each forward model starting from Case 1 to Case 4, respectively. The first column depicts retrieval of CH$_4$, the second column is for NH$_3$ and the last column is for H$_2$O. The red dotted line shows the input mixing ratio profile. The shaded blue region shows the posterior for the retrieval in each case. {\bf Main Point}- Case 3 and 4 retrieve molecular mixing ratios accurately, but not always precisely (i.e. NH$_3$).}
\label{fig:fig5}
\end{figure}

%fig5=fig 10
Figure \ref{fig:fig5} presents the comparison of the retrieved molecular mixing ratios compared with the input molecular mixing ratio profiles, which was used to generate the simulated observed spectra for {\it 47 Uma b}. In this first case, a single value for the mixing ratio was retrieved. The full posteriors of this single retrieved value are depicted in light blue, while the altitude-dependent profile is shown in dashed red. 

H$_2$O is the most dominant molecular opacity source for {\it 47 Uma b}, among the three molecules. The H$_2$O opacity causes broadband molecular absorption features starting from 0.5-1$\mu$m. Constraints on the H$_2$O mixing ratio for Case 1 and 2 are precise but not accurate. Constraints on H$_2$O mixing ratio by Case 3 and Case 4 are relatively accurate, but the precision appears to be obscured by the altitude-dependence. %This is not very intuitive from the seemingly featureless spectra that has been modeled for {\it 47 Uma b}. 
%Additionally, the H$_2$O input profile shows a depletion in water abundance above 10$^{-3}$ bars because of the condensation. In these simplified cases, we only retrieve a constant mixing ratio profile. 
The posterior of the H$_2$O mixing ratios by Case 3 and 4 is most highly peaked at the region of highest water mixing ratio of the `true' atmosphere state (deep in the atmosphere). However, the posterior contains an additional tail that extends to the depleted value. This might indicate that there is sensitivity to the depleted abundances above the cloud deck. In \S \ref{sec:depletedwater} we report the results of adding in an additional parameter to capture the depleted water profile.

%%%%%%%%%%% THIS FIGURE HAS BEEN UPDATED WITH NEW RUNS

CH$_4$ is retrieved by both Case 3 and Case 4 within 2$\sigma$ of the `true' CH$_4$ profile, though the magnitude of the 2$\sigma$ value is relatively large (an order of magnitude in abundance), and the posterior is not Gaussian. For this planet-case, this result is not surprising. CH$_4$ contributes to some of the absorption features beyond 0.7~$\mu$m but unlike \textit{eps Eri b}, the CH$_4$ opacity contribution is smaller than H$_2$O at all wavelengths. While Case 3 retrieves CH$_4$ well within the 2$\sigma$ limit of the input CH$_4$ mixing ratio profile, Case 4 better constrains CH$_4$ within 1$\sigma$. Like H$_2$O, CH$_4$ is overestimated by Case 1 and 2 with sharply peaked posteriors. %Case 1 still retrieves the CH$_4$ within the 2$\sigma$ limit, which is comparatively better than the Case 2 retrieval. 

In all cases, SNR$=$20 does not allow for the precise retrieval of an NH$_3$ abundance because, although it is relatively abundant, its opacity contribution to the spectrum is negligible until 0.8$\mu$m. %NH$_3$ opacity contribution peaks very sharply past 0.8 $\mu$m.
Case 1 and 2 fail to constrain NH$_3$ at all and return the produce nearly uniform posteriors (the prior). Case 3 also returns a posterior whose $1\sigma$ constraint spans $\sim$2 orders of magnitude. %Case 3 constrains NH$_3$ just within 1$\sigma$ but the posterior is relatively broad causing the retrieval to be less precise compared to the other two molecules retrieved by Case 3. 
The retrieved posterior distribution for NH$_3$ with Case 4 shows a bias towards higher NH$_3$ abundance, and does not yield a Gaussian posterior, which points to some degeneracy. With the CH$_4$, H$_2$O and NH$_3$ retrievals for {\it 47 Uma b} described above, it is clear that Case 1 and 2 are not robust models for retrieving abundances while Case 3 and 4 retrieve the mixing ratios of nearly all the three molecules within 2$\sigma$ of the true profiles.  

\begin{figure}
  \centering
  \includegraphics[width=0.5\textwidth]{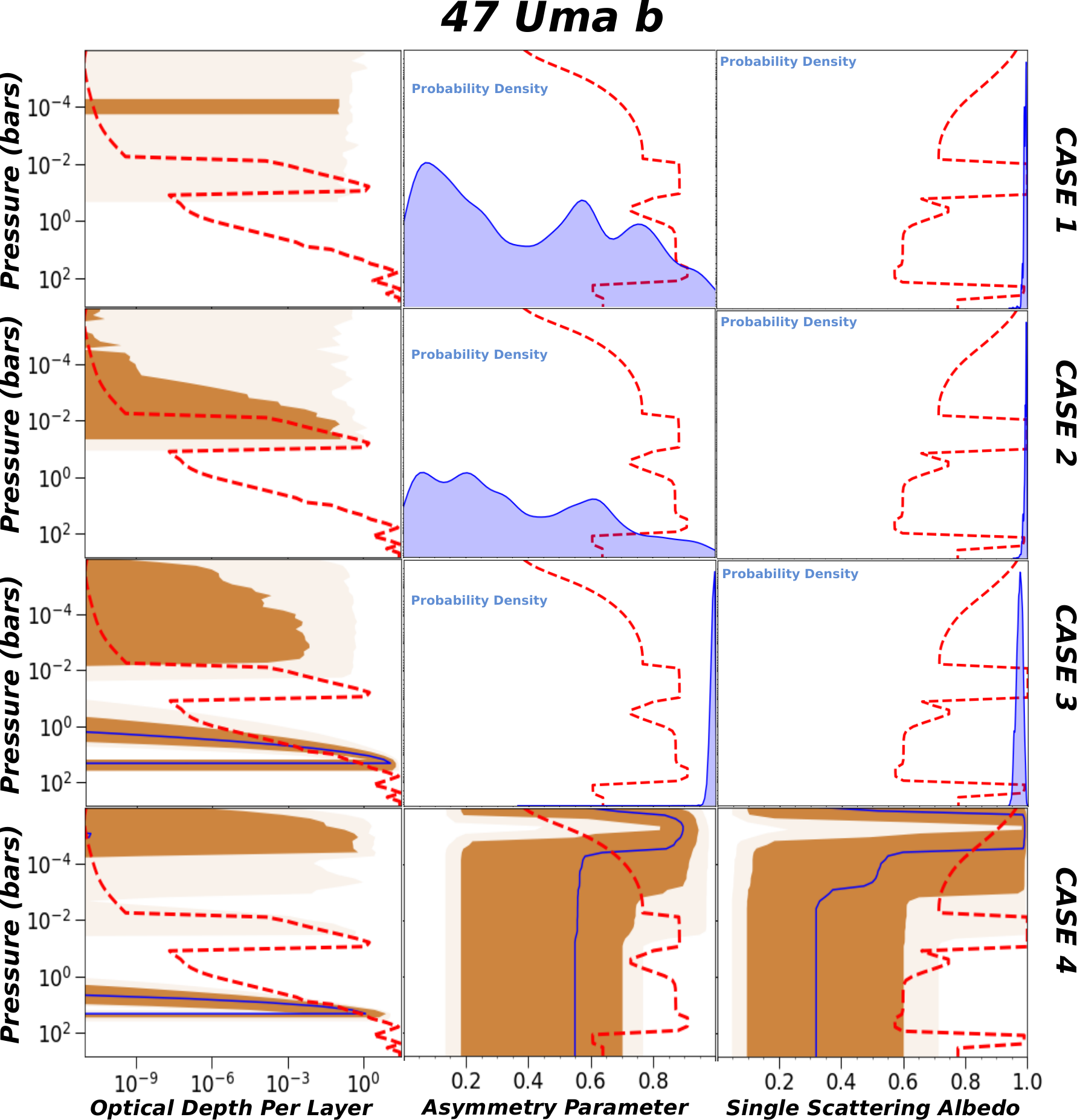}
  
  \caption{This figure shows retrieved cloud structure and optical properties compared with the input cloud structure for {\it 47 Uma b}. The four rows depict the retrievals for each parameterization starting from Case 1 to Case 4, respectively. The first column shows the optical depth per layer, the second column is the asymmetry parameter and the last column is the single scattering albedo. The red dashed line shows the input cloud structure averaged over the wavelength range of 0.3 to 1. The dark and light brown colored patches depict the 1$\sigma$ and 2$\sigma$ confidence intervals for those cases where altitude-dependence was considered. The blue shaded curves depict the posterior distribution for a singly retrieved parameter. %{\bf Main Point}- Each of the four cases perform differently in retrieving cloud properties.
  }
\label{fig:fig6}
\end{figure}

Cloud retrievals for each case are shown in Figure \ref{fig:fig6}. The dashed red lines show the 0.3-1$\mu$m averaged input profiles produced by \texttt{Virga}.  The left column shows the optical depth per layer retrievals for the four cases. The middle and right column shows the retrieval of the asymmetry and single scattering albedo, respectively. The median profiles (in blue) along with 1$\sigma$ and 2$\sigma$ bounds (in dark and light brown) are shown for those cases where altitude-dependence was considered. For the rest, the posteriors for the single parameter are shown in with the blue shaded curves. 

%%%%%%%%%%% THIS FIGURE HAS BEEN UPDATED WITH NEW RUNS

The optical depth per layer profiles show that %the cloud pressure of the upper cloud deck is retrieved accurately by Case 2. 
both the single deck parameterizations -- Case 1 and 2 -- trace only the upper cloud deck. Case 1 retrieves the top H$_2$O cloud deck position more precisely compared to Case 2. Whereas Case 2 retrieves the top cloud deck position more accurately compared to Case 1. Case 3 retrieves the bottom cloud deck position and optical depth profile accurately but fails to retrieve accurate or precise parameters for the top cloud deck. Similarly, Case 4 gets the bottom deck but places the top cloud deck at much lower pressures ($<10^{-4}$bars) missing the `true' cloud deck pressure by several ($\sim$ 4) orders of magnitude. 

The retrieved asymmetry parameter by Case 1 and 2 are highly degenerate (the posterior is multi-modal). Case 3 retrieval of the asymmetry parameter profile is relatively precise but it overestimates the asymmetry parameter value. Case 4 retrieves an inaccurate asymmetry parameter profile with a low ($\sim$ 0.2) asymmetry parameter across most of the atmosphere except a large value ($\sim$ 0.9) at the position of the retrieved top cloud deck. 

High values ($\sim$ 1) of the single scattering albedo are retrieved by Cases 1,2 and 3. These three retrieval models hence prefer reflective cloud particles, similar to H$_2$O. This is not the case for Case 4, where the retrieved bottom deck is relatively unreflective with a single scattering albedo of 0.4, whereas the top deck is highly reflective with single scattering albedo close to 1. 

None of the four retrieval models succeed in retrieving the cloud structure for {\it 47 Uma b} completely. Some aspect of the position of the cloud layers are accurately captured by Case 2, 3 and 4. For example, Case 3 and 4 retrieve the position and the optical depth profile of the deeper bottom cloud deck accurately, but incorrectly retrieve either the position or optical depth of the top layer. We discuss this further in \S \ref{sec:discussion}.  %The cloud and molecular retrievals with the four cases for {\it 47 Uma b} show that Case 3 and 4 clearly outperform the other cases in retrieving the cloud structure and the molecular abundances for this planet. 
 %Next, we explore the sensitivity of Case 3 retrieval to the depleted H$_2$O profile of {\it 47 Uma b}.

\subsubsection{47 Uma b Case 3 with a Depleted Water Profile}{\label{sec:depletedwater}}
The posterior distribution of the well-mixed H$_2$O mixing ratio retrieved using Case 3 on {\it 47 Uma b} shows a bias towards lower abundances of H$_2$O, while the retrieved median value coincides with the H$_2$O mixing ratio below the cloud deck in the `true' atmosphere. This bias, shown in Figure \ref{fig:fig5}, might be indicative of model sensitivity with respect to the depletion above the top H$_2$O cloud deck. This sensitivity was similarly highlighted in the work of \citet{Damiano2020exorel}.  Here, we modify the well-mixed H$_2$O retrieval model in Case 3 to accommodate a depleted water mixing ratio above the top cloud deck. This requires two additional parameter for Case 3 since we now retrieve two H$_2$O mixing ratios -- one below (H$_2$O$^1$) and one above (H$_2$O$^2$) the cloud deck. In the retrieval model, we define the cloud top pressure, P$_{\rm top}$, as the location where the optical depth of the top cloud deck is 10$^z$ times less than the peak cloud optical depth, where $z$ is the second additional parameter. The water mixing ratio is modeled to be a straight line with negative slope in the log(pressure) vs. log(H$_2$O) space between the lower and upper cloud deck pressure. 

%%%%%%%%%%% THIS FIGURE HAS BEEN UPDATED WITH NEW RUNS
\begin{figure}
  \centering
  \includegraphics[width=0.45\textwidth]{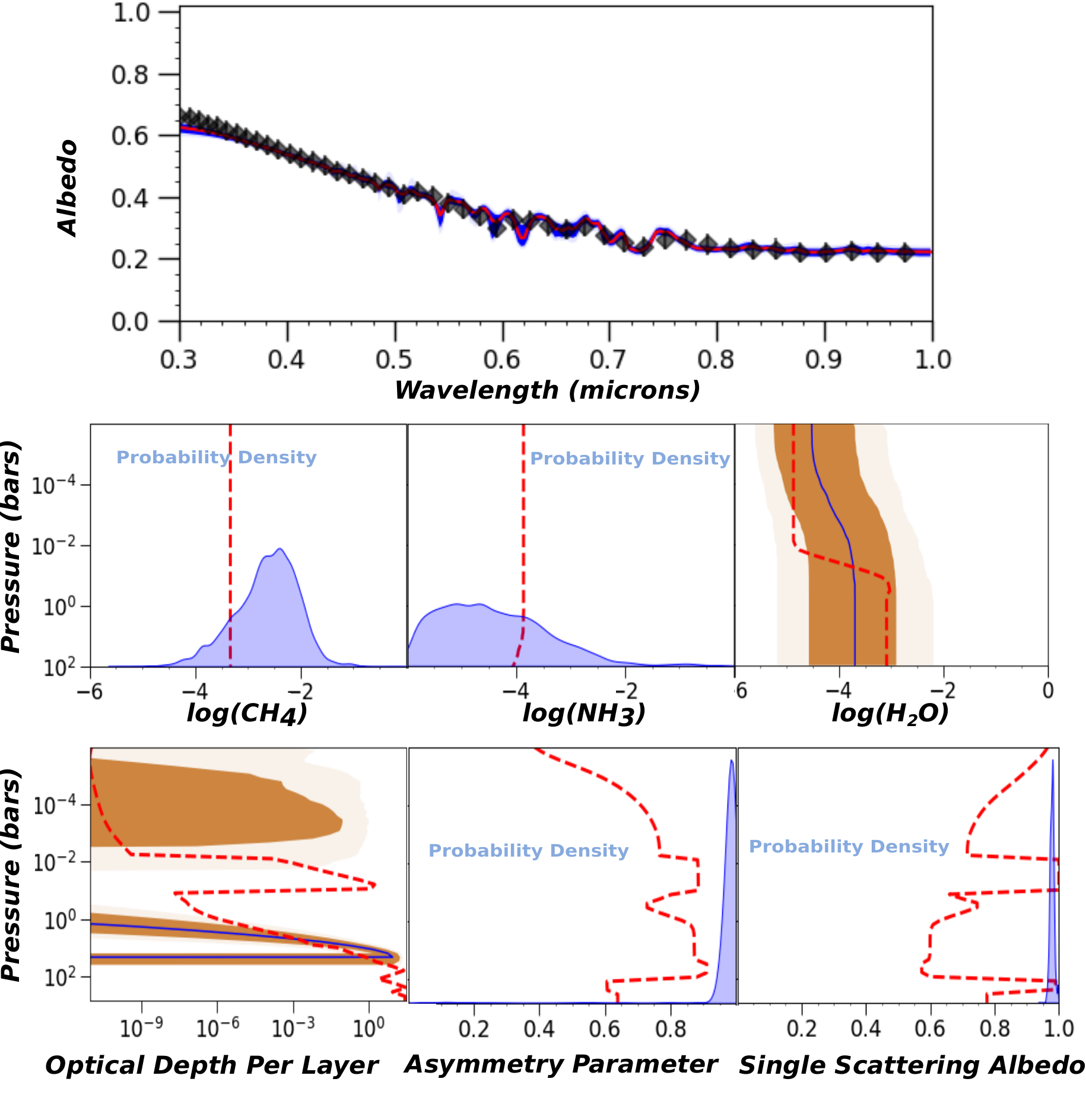}
  
  \caption{Comparison of the retrieved atmospheric properties of {\it 47 Uma b} using the modified Case 3 model accounting for H$_2$O depletion above the cloud deck. The top row shows the comparison of the observed spectra in black diamonds with the median (red), 1$\sigma$ (dark blue) and 2$\sigma$(light blue)  regions of the retrieved spectra. The middle row shows the retrieved mixing ratios of CH$_4$, NH$_3$ and H$_2$O from left to right with the true input profiles shown in red. The last row shows the retrieved cloud properties -- optical depth per layer, asymmetry parameter and single scattering albedo from left to right with the true input profiles shown in red. {\bf Main Point}- At this SNR=20, retrieving a depleted water profile above the cloud deck for a {\it 47 Uma b}-like planet can lead to an improvement in the constraint interval of cloud and abundance properties}
\label{fig:dep}
\end{figure}

Figure \ref{fig:dep} shows the spectral fit of the retrieved results along with the cloud and molecular abundance retrievals.
We find that this additional complexity within Case 3 doesn't improve the retrieved cloud optical depth structure significantly for {\it 47 Uma b} compared to the well-mixed Case 3 retrieval. Without the depleted water profile, Case 3 is unable to constrain the top cloud deck pressure within the lower bounds of the cloud deck pressure parameter space but with the depleted water profile, the additional complexity in this modified Case 3 model helps to put 1$\sigma$ bounds on the top cloud pressure within the lowest pressure bound of the retrieved atmospheric model. The retrieval of the deeper cloud deck remains similarly accurate/precise in both the models. This parameterization also broadens the constraints on the asymmetry parameter towards lower values but the overestimation of the asymmetry parameter as seen for Case 3 in Figure \ref{fig:fig6} still persists. Retrieval of the single scattering albedo remains similar in both the cases.

This model captures the depleted H$_2$O abundance above the cloud deck within 1$\sigma$ and can also retrieve the deep H$_2$O abundance within 1$\sigma$. This modified model does not show any improvement in the retrieval of CH$_4$ over the original Case 3 model. NH$_3$ is better constrained by the modified Case 3 model compared to Case 3 (unmodified) where NH$_3$ is not at all constrained.

Despite the apparent improvement in water abundance, a Bayes factor analysis suggests that Case 3 \textit{without} water depletion is weakly preferred over Case 3 with water depletion for {\it 47 Uma b}. The Bayes factor difference is so small that without prior knowledge of the true solution, it would be difficult to discern which scenario was correct. That is, the complexity that arises from the addition of the depleted profile parameter, is not strongly favored.

\subsubsection{Retrieval at Higher Spectral Resolution (R $=$ 140)}

 Future mission concepts like LUVOIR and HabEx are expected to achieve a spectral resolution of $\sim$ 140 in optical wavelengths as shown in Table \ref{table:0}. We retrieve on a R $\sim$ 140 and SNR $\sim$ 20 spectra of {\it 47 Uma b} with %the single cloud deck model, Case 2, and 
the double cloud deck model, Case 3, to investigate how a higher spectral resolution from these missions can improve upon the constraints on the molecular abundances and cloud properties obtained from lower spectral resolution data expected from {\it Roman Space Telescope}.

\begin{figure*}
  \centering
  \includegraphics[width=1\textwidth]{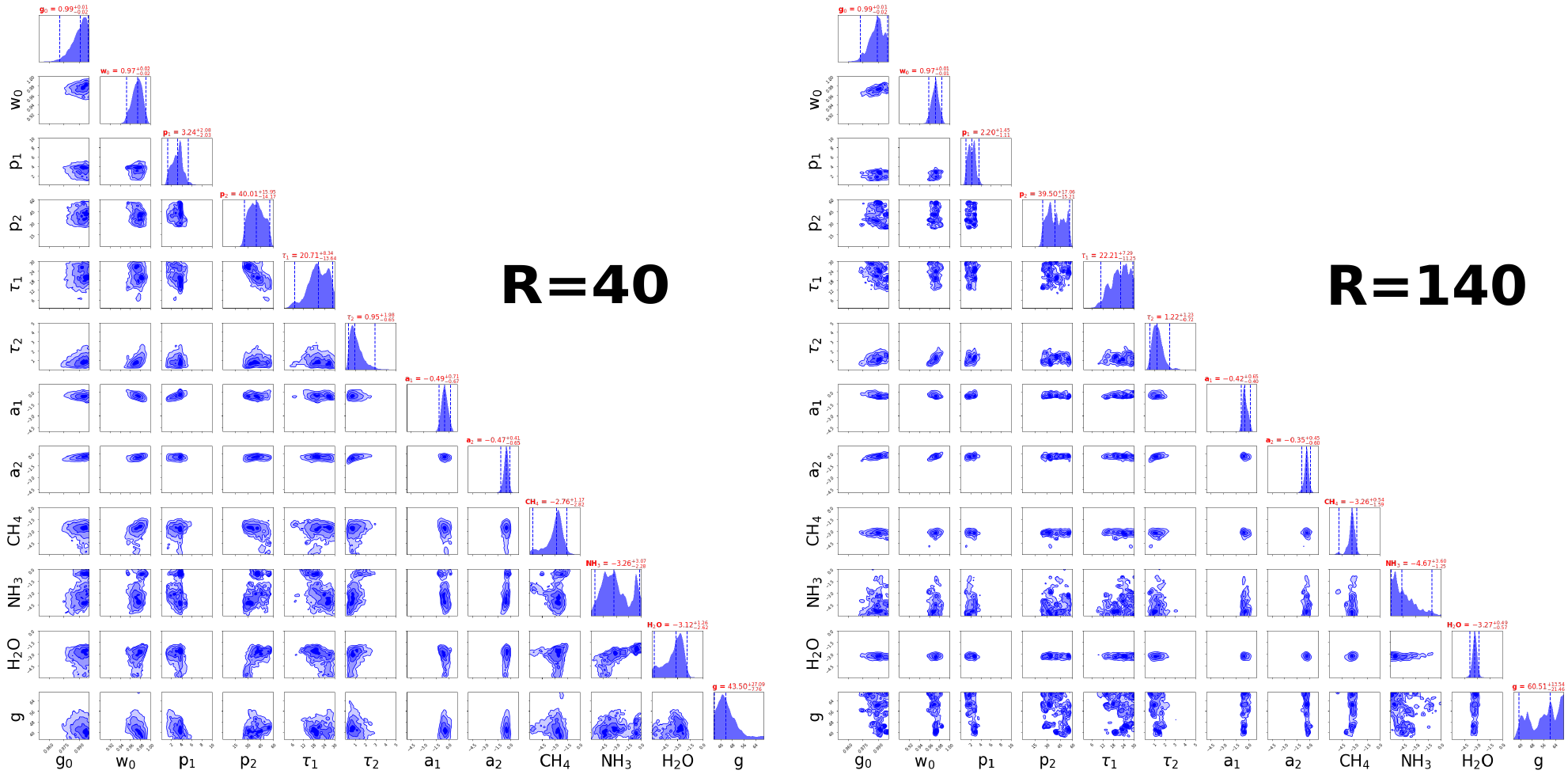}
  \caption{ Corner Plot of the Case 3 retrieval on the {\it 47 Uma b} reflected spectra with R $\sim$ 40 (left panel) and R $\sim$ 140 (right panel). This shows the posteriors of the parameters described in Table \ref{table:3} and correlations between them. The median and the 2$\sigma$ confidence intervals on each parameter is shown with blue vertical lines in the posterior distribution panel of each parameter. The median value of each parameter along with their 2$\sigma$ errors are also reported for each parameter. \textbf{Main point-} Although higher resolution improves the accuracy and precision of molecular abundances, it does not largely improve that of the cloud parameters.}
  
\label{fig:fignew}
\end{figure*}

Figure \ref{fig:fignew} left and right panel shows the corner plots of the retrieved Case 3 parameters from the {\it 47 Uma b} reflection spectra with spectral resolution of R $\sim$ 40 and 140, respectively. Here, we opt to show full set of marginalised posterior distributions of each parameter, in order to showcase correlations between the various retrieved parameters. Overall, higher resolution spectra leads to tighter constraints on multiple parameters, relative to the R $\sim$ 40 spectra with Case 3. 

%Specifically, higher spectral resolution leads to significantly better constraint on molecular mixing ratios of both CH$_4$ and H$_2$O. 
Specifically, the bias in the H$_2$O retrieval seen in the lower resolution retrieval with Case 3 is no longer present in the posterior retrieved from the the higher resolution spectra. The 2$\sigma$ constraints on both H$_2$O and CH$_4$ has tightened by a factor of $\sim$2 relative to the posteriors shown in Figure \ref{fig:fignew}. CH$_4$ retrieval has improved from being within 2$\sigma$ of the `true' CH$_4$ abundance in the low resolution case to being within 1$\sigma$ at high resolution. The NH$_3$ posteriors remain unconstrained in both spectral resolutions -- though this is as expected given it's opacity contribution at these wavelengths. Lastly, although the abundances improve significantly, retrievals of the cloud parameters remain very similar (within 1$\sigma$) to that obtained from the lower resolution spectra. This further motivates the utility of low resolution spectroscopy for obtaining cloud properties.

%The Case 2 retrieval model overestimates the molecular abundances of both CH$_4$ and H$_2$O even in this high resolution retrieval as was seen in Figure \ref{fig:fig5}. NH$_3$ abundance remains unconstrained with case 2. So, for {\it 47 Uma b} like planets single cloud deck models like case 2 are not reliable even with higher spectral resolution data. 

%\begin{figure*}
%  \centering
%  \includegraphics[width=0.45\textwidth]{47umab_70_case3.pdf}
%  \includegraphics[width=0.45\textwidth]{47umab_70_case3.pdf}
  
%  \caption{Comparison of the retrieved atmospheric properties of {\it 47 Uma b} when higher resolution (R $\sim$ 140) reflection spectra is retrieved on. The left panel shows the retrieval results using Case 2 parameterization while the right column shows retrieval results with Case 3 parameterizations. The top row shows the comparison of the observed spectra in black diamonds with the median (red), 1$\sigma$ (dark blue) and 2$\sigma$ (light blue) regions of the retrieved spectra with each parameterization. The middle row shows the retrieved mixing ratios of CH$_4$, NH$_3$ and H$_2$O from left to right in both the columns with the true input profiles shown in red. The last row shows the retrieved cloud properties -- optical depth per layer, asymmetry parameter and single scattering albedo from left to right in both the columns with the true input profiles shown in red.}
%\label{fig:highres}
%\end{figure*}

\subsection{ eps Eri b}

\begin{figure*}
  \centering
  \includegraphics[width=0.45\textwidth]{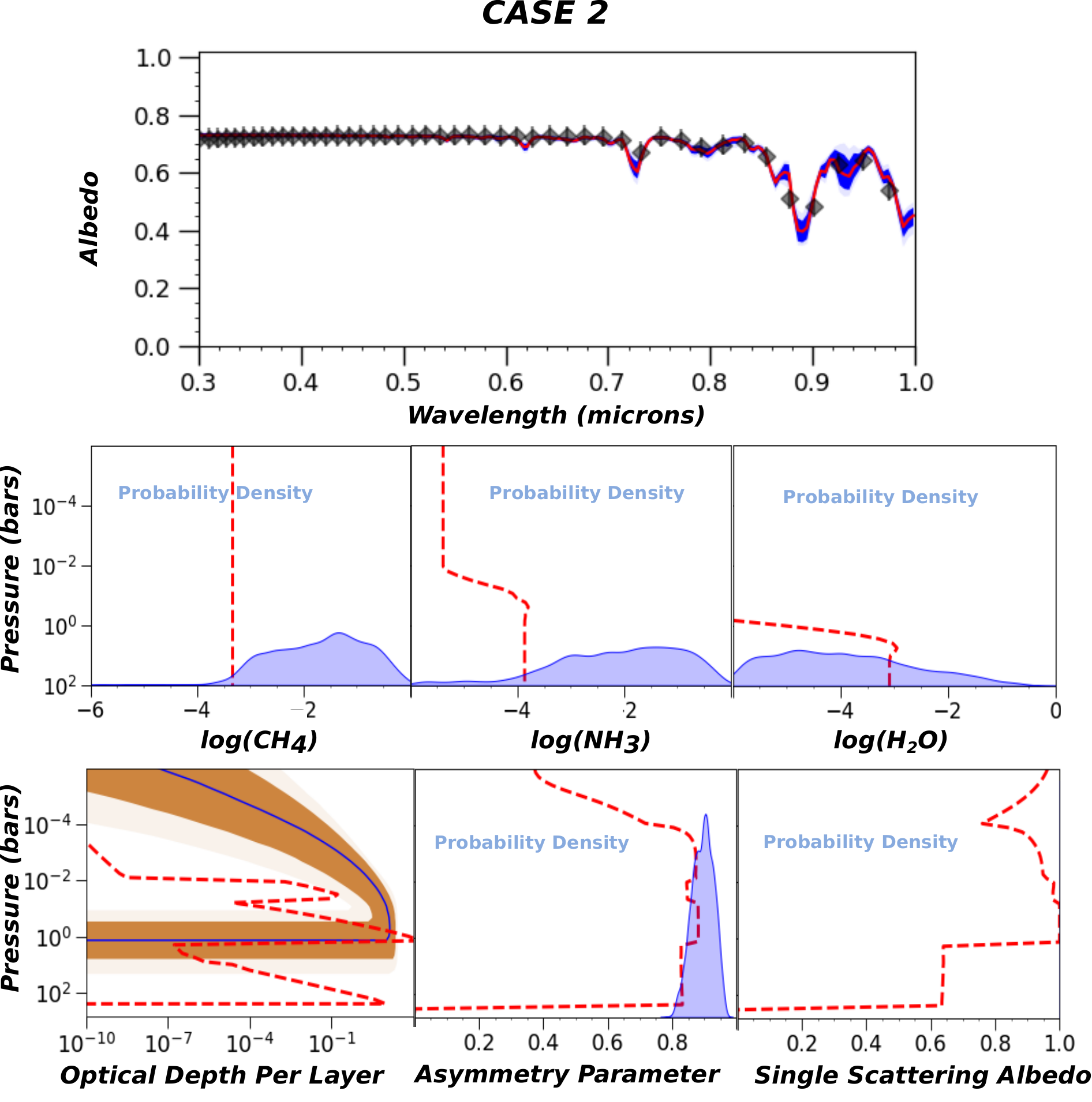}
  \includegraphics[width=0.45\textwidth]{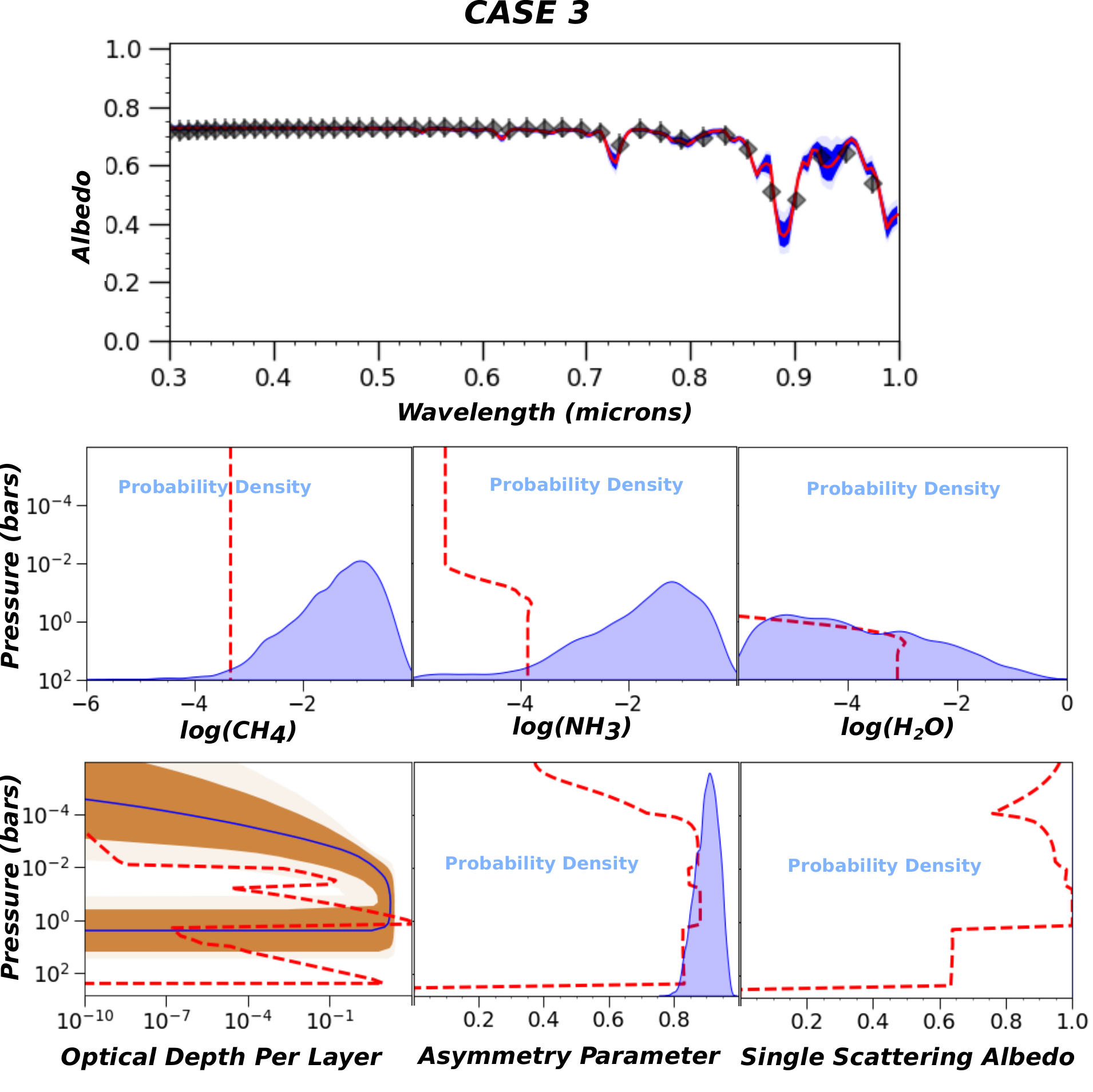}
  
  \caption{Comparison of the retrieved atmospheric properties of {\it eps Eri b}. The left panel shows the retrieval results using Case 2 parameterization while the right column shows retrieval results with Case 3 parameterizations. The top row shows the comparison of the observed spectra in black diamonds with the median (red), 1$\sigma$ (dark blue) and 2$\sigma$ (light blue) regions of the retrieved spectra with each parameterization. The middle row shows the retrieved mixing ratios of CH$_4$, NH$_3$ and H$_2$O from left to right in both the columns with the true input profiles shown in red. The last row shows the retrieved cloud properties -- optical depth per layer, asymmetry parameter and single scattering albedo from left to right in both the columns with the true input profiles shown in red. {\bf Main Point}- Case 2 performs slightly better than Case 3, and is preferred according to Bayes factor.}
\label{fig:eps_cases}
\end{figure*}

Out of our 3 planet sample, the spectroscopic model of {\it eps Eri b} is the most similar to Jupiter, and it is our coolest target. The reflected spectra is dominated by cloud opacity, as seen in Figure \ref{fig:figphot}. Therefore, it has a zero-sloped reflection spectra with a major CH$_4$ feature at 0.73 \& 0.9$\mu$m along with a minor NH$_3$ feature. The retrieval results are shown in \ref{fig:eps_cases}. 

Overall, the accuracy and precision of the retrieved cloud and mixing ratio profiles of the atmosphere were found to be very similar for Case 2 and 3, which negates the addition of more complexity in going to Case 4. Moreover, the Bayes factor calculation suggests Case 2 is weakly favored over Case 3.

Despite the multiple cloud decks made up of different condensate species, Case 3 retrieves a single deck. The maximum optical depth of the highest pressure water cloud located at 1 bar is $\tau>10$. The NH$_3$ cloud directly above the water deck only reaches a maximum optical depth of $\tau>0.5$. Therefore, when retrieving the optical depth profile, the model favors a single larger cloud deck, which spans both the H$_2$O and the NH$_3$ cloud. Despite H$_2$O and the NH$_3$ having different optical properties, the asymmetry parameter and single scattering albedo structure for the planet is retrieved accurately and precisely within the 1$\sigma$ bounds of the input profile by both the cases. Here, we define accuracy as retrieving the true value in the region of maximum cloud opacity. 

Both Case 2 and 3 overestimate the CH$_4$ abundance for {\it eps Eri b} %. The CH$_4$ abundance estimate from Case 2 is marginally better than the estimate from Case 3. However, the CH$_4$ is overestimated 
despite the presence of two prominent CH$_4$ features in the {\it eps Eri b} albedo spectra. This CH$_4$ overestimation was also noted in \citet{lupu2016developing} for the case of {\it HD 99492 c}, which also contained two CH$_4$ absorption features similar to the {\it eps Eri b} spectra in our case. NH$_3$ is neither precise nor accurate, which is intuitive given the opacity contribution to the spectrum. H$_2$O remains unconstrained for {\it eps Eri b} because the `true' H$_2$O mixing ratio depletes at high pressures ($>1$~bars) due to water cloud formation. Therefore, there is a limited H$_2$O contribution to the albedo spectra for {\it eps Eri b}. 

We performed additional retrieval tests for {\it eps Eri b} in order to determine the source of the overestimated CH$_4$. The tests we performed included: 1) fixing the gravity to the true value, 2) increasing the resolution of the simulated spectrum to R=120, and 3) removing water as a free parameter. Fixing gravity, and removing of H$_2$O did not increase the accuracy or precision of the CH$_4$ abundance. Increasing the resolution to R=120 increased the accuracy of the retrieved CH$_4$ by a two orders of magnitude, which is a clear indication that higher resolution is needed to accurately retrieve CH$_4$. At R=120, 4 data points sample the under-saturated CH$_4$ feature at 0.7 microns, while 8 data points sample the saturated feature at 0.9 microns. This is in contrast to R=40 where only 1 data point samples the under-saturated feature at 0.7 microns and 2 data points sample the feature at 0.9 microns. 

%%%%%%%%%%% THIS FIGURE HAS BEEN UPDATED WITH NEW RUNS
%We discuss this further in \S \ref{sec:discussion}.

\subsection{HD 62509 b}

{\it HD 62509 b } is our hottest target planet with an estimated effective temperature of $\sim$533 K. The clouds in this planet are formed much deeper in the atmosphere compared to the other two cooler targets in our consideration, as is evident in Figure \ref{fig:figtp}. As a result, most of the spectra is dominated by molecular opacity and Rayleigh scattering as shown in Figure \ref{fig:figphot}. Additionally, the planet is dim compared to the other two planets in reflected light. Similar to {\it eps Eri b}, the Bayes factor calculation suggests Case 2 is weakly preferred over Case 3 for {\it HD 62509 b }. Therefore, the accuracy and precision of both Case 2 and Case 3 in retrieving cloud properties and the molecular abundances are very similar for {\it HD 62509 b}.

\begin{figure*}
  \centering
  \includegraphics[width=0.45\textwidth]{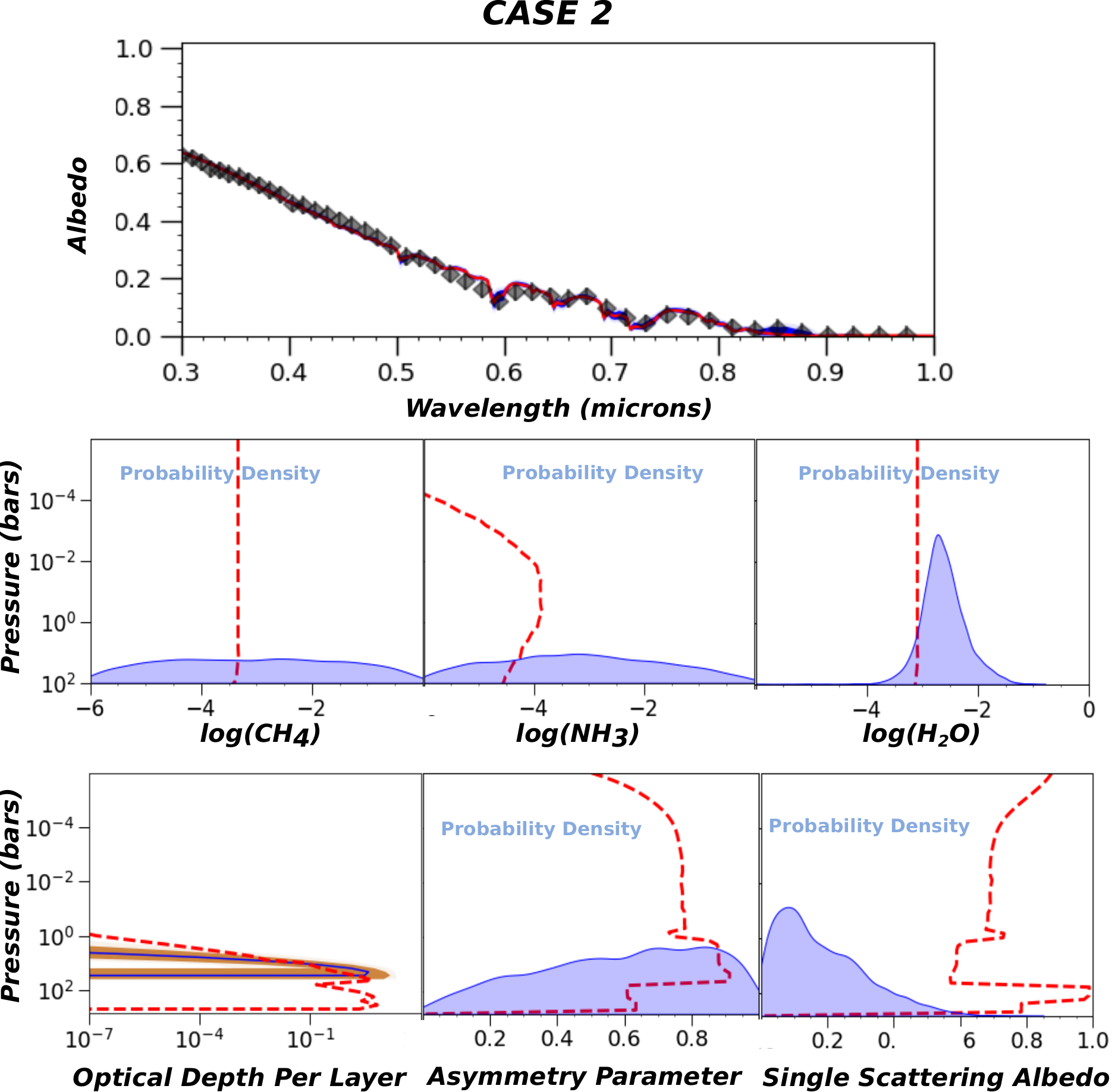}
  \includegraphics[width=0.45\textwidth]{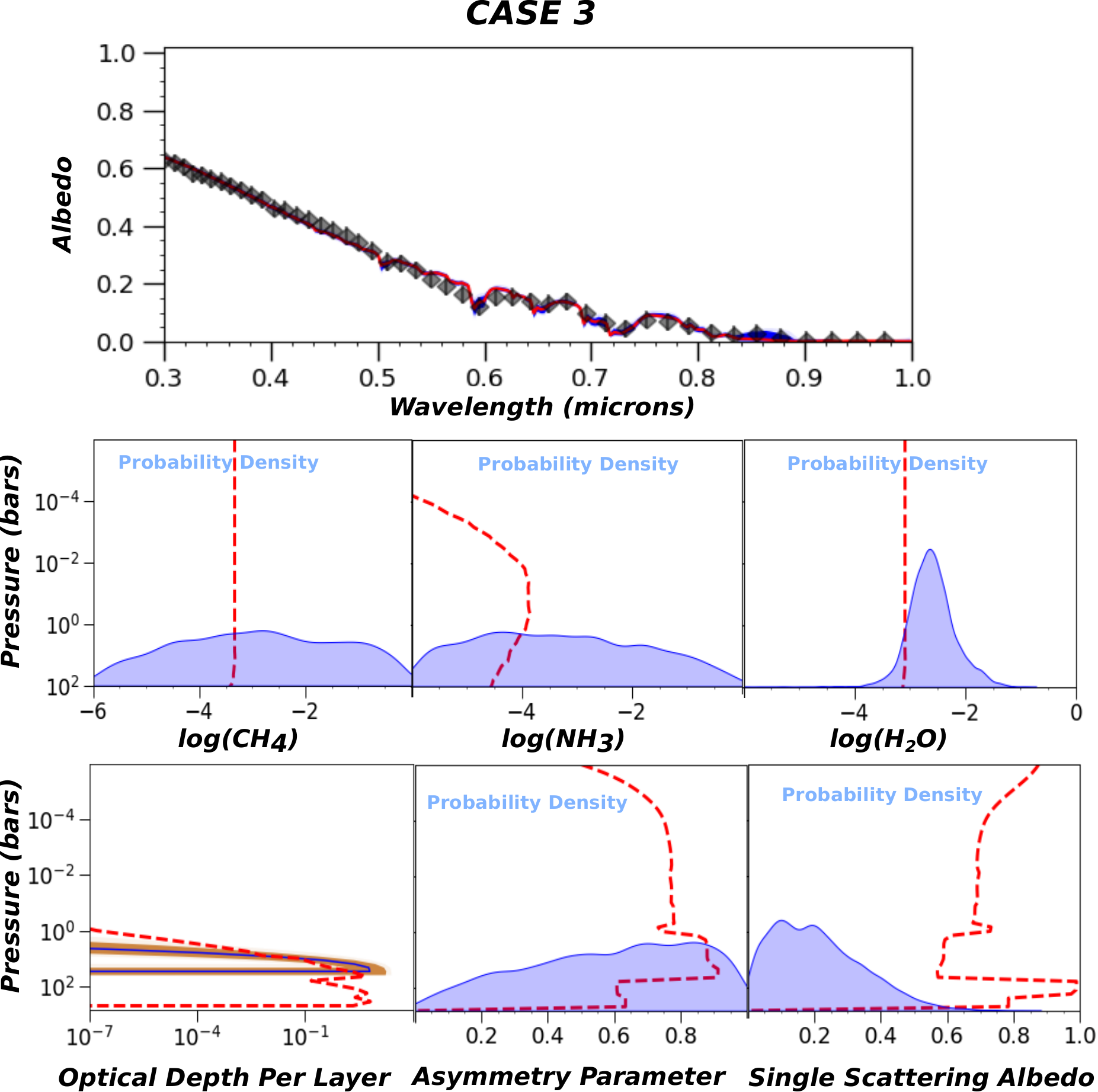}
  
  \caption{Comparison of the retrieved atmospheric properties of {\it HD 62509 b}. The left panel shows the retrieval results using Case 2 parameterization while the right column shows retrieval results with Case 3 parameterizations. The top row shows the comparison of the observed spectra in black diamonds with the median (red), 1$\sigma$ (dark blue) and 2$\sigma$ (light blue) regions of the retrieved spectra with each parameterization. The middle row shows the retrieved mixing ratios of CH$_4$, NH$_3$ and H$_2$O from left to right in both the columns with the true input profiles shown in red. The last row shows the retrieved cloud properties -- optical depth per layer, asymmetry parameter and single scattering albedo from left to right in both the columns with the true input profiles shown in red. {\bf Main Point}- Case 2 performs slightly better than Case 3 but the retrieved results are extremely similar.}
\label{fig:hd_cases}
\end{figure*}

H$_2$O mixing ratio is retrieved accurately within 2$\sigma$ by Case 2 \& 3 for {\it HD 62509 b} as shown in Figure \ref{fig:hd_cases}. This is because, like {\it 47 Uma b}, H$_2$O is the most significant gaseous opacity source in the atmospheres of {\it HD 62509 b}. Both Case 2 \& 3 fail to constrain the CH$_4$ and NH$_3$ abundance for {\it HD 62509 b} since the spectra is dominated by H$_2$O opacity for {\it HD 62509 b} with negligible opacity contribution from CH$_4$ and NH$_3$.

For {\it HD 62509 b}, the Case 2 \& 3 model traces the region where the lower cloud deck achieves optical depth of $\sim$10. The retrieval is able to constrain the cloud optical depth structure despite: 1) the cloud being very deep in the atmosphere and 2) the relatively small cloud opacity contribution to the reflected spectra compared to our two other target planets. Both Case 2 and 3 underestimate the single scattering albedo for {\it HD 62509 b}. This is likely because the cloud deck is at such depth in the atmosphere that molecular absorption dominates the total opacity. The retrieval models hence become less sensitive to the reflectivity of the cloud particles. The posterior of the asymmetry parameter parameter nearly spans the entire prior region. This too is because of the lack of sensitivity of the model to the cloud scattering properties due to the depth of the cloud in the atmosphere. 

%%%%%%%%%%% THIS FIGURE HAS BEEN UPDATED WITH NEW RUNS

\section{Discussion}\label{sec:discussion}

\subsection{Required Cloud Complexity}% Depends on Relative Location of Molecular and Cloud Opacity}

Between the first two cases with 9 parameters, Case 1 (box model) could never accurately/precisely retrieve atmospheric or scattering properties. Therefore, we advocate against box models for future retrieval work, even in simplified studies. 

Between Cases 2 and Case 3 (from one to two cloud decks), the results were less clearly defined. The reflected light spectrum of {\it eps Eri b} is dominated by cloud opacity, while the hotter {\it HD 62509 b } spectrum is dominated by molecular and Rayleigh opacity. In both the cases we find that a single deck model (e.g. Case 2) can sufficiently capture the atmospheric properties and produce similar results to the double cloud deck model (e.g. Case 3). For the case of {\it 47 Uma b}, though, double deck models (Case 3 \& 4) are significantly favored over single deck models (Case 1 \& 2). Naively, this is contrary to the intuition that might be gained from looking at Figure \ref{fig:figcldopd}, which shows a strong NH$_3$ \& CH$_4$ double cloud deck for {\it eps Eri b}.

The key factor that dictated whether an additional cloud deck was necessary was the pressure location of the highest optical depth ($\tau\sim10$) cloud deck --  relative to the region of maximum contribution of molecular opacity. The {\it eps Eri b} NH$_3$ cloud deck achieved $\tau\sim0.5$ at 0.5~bars. The H$_2$O deck achieved $\tau\sim10$ at 1~bar. The bulk of the molecular opacity (see Figure \ref{fig:figphot}) was mostly contained below this optically thick H$_2$O deck (e.g. cloud layer-cloud layer-molecular opacity). This is in contrast to {\it 47 Uma b}, where the upper cloud deck achieved $\tau\sim1$ at 0.1~bars while lower deck did not achieve $\tau\sim10$ until roughly 30~bars. The molecular opacity sat between these two regions (e.g. cloud-molecular-cloud). Therefore, the former case of {\it eps Eri b} could be accurately modeled with one larger cloud deck, where as the latter case of {\it 47 Uma b} required two separate scattering regions both below and above the region of highest molecular opacity. Of course we will not know \textit{a priori} where these cloud decks exist. However, this result can inform the interpretation of future reflected light results. Retrievals that strongly favor single deck models may not ultimately reflect the true state. 

%Lastly, Case 4 -- two parameters to describe asymmetry and single scattering albedo -- was {\bf the most} favored model according to Bayes factor analysis. This is because precision with which single scattering albedo and asymmetry were constrained ($>\pm0.1$) was larger than the total variation in the profiles in the region of high cloud opacity. Therefore, Case 4 might still be favorable for cases not explored here. For example, a case with a water cloud deck below high-altitude photochemically-produced hazes \citep[e.g.][]{gao2017sulfur} might require at least 2-parameters in each of the scattering properties.  

 Lastly, Case 4 -- two parameters to describe asymmetry and single scattering albedo -- was only moderately favored over Case 3, according to Bayes factor analysis. However, Case 4 failed to retrieve an accurate and precise single scattering albedo and the asymmetry parameter profile for {\it 47 Uma b}. %even though it had additional complexity in the parametrizations of these two scattering property profiles specifically compared to all the other three cases.
Additionally, the accuracy of the retrieved molecular abundances with Case 4 were also similar to Case 3 (within 1$\sigma$). Even though the use of Case 4 was not strongly motivated in this work, it could be useful for cases not explored here. Specifically, the use of Case 4 would be suitable for atmospheres that have stronger vertical variation in asymmetry, or single scattering albedo. For example, a case with a water cloud deck below high-altitude photochemically-produced hazes \citep[e.g.][]{gao2017sulfur} might require at least 2-parameters in each of the scattering properties.

\subsubsection{Ability to Constrain Gravity}
Unlike for most transiting planets, there generally will only be approximate constraints on gravity for directly imaged planets in reflected light. While radial velocity and a sufficient number of images will constrain $\sin i$, planet radii will still be uncertain. Therefore it is worthwhile to determine 1) whether or not gravity can be accurately retrieved from reflected light spectroscopy alone, and 2) whether or not an imprecise gravity effects the ability to retrieve accurate atmospheric properties. 

\begin{figure}
  \centering
  \includegraphics[width=0.5\textwidth]{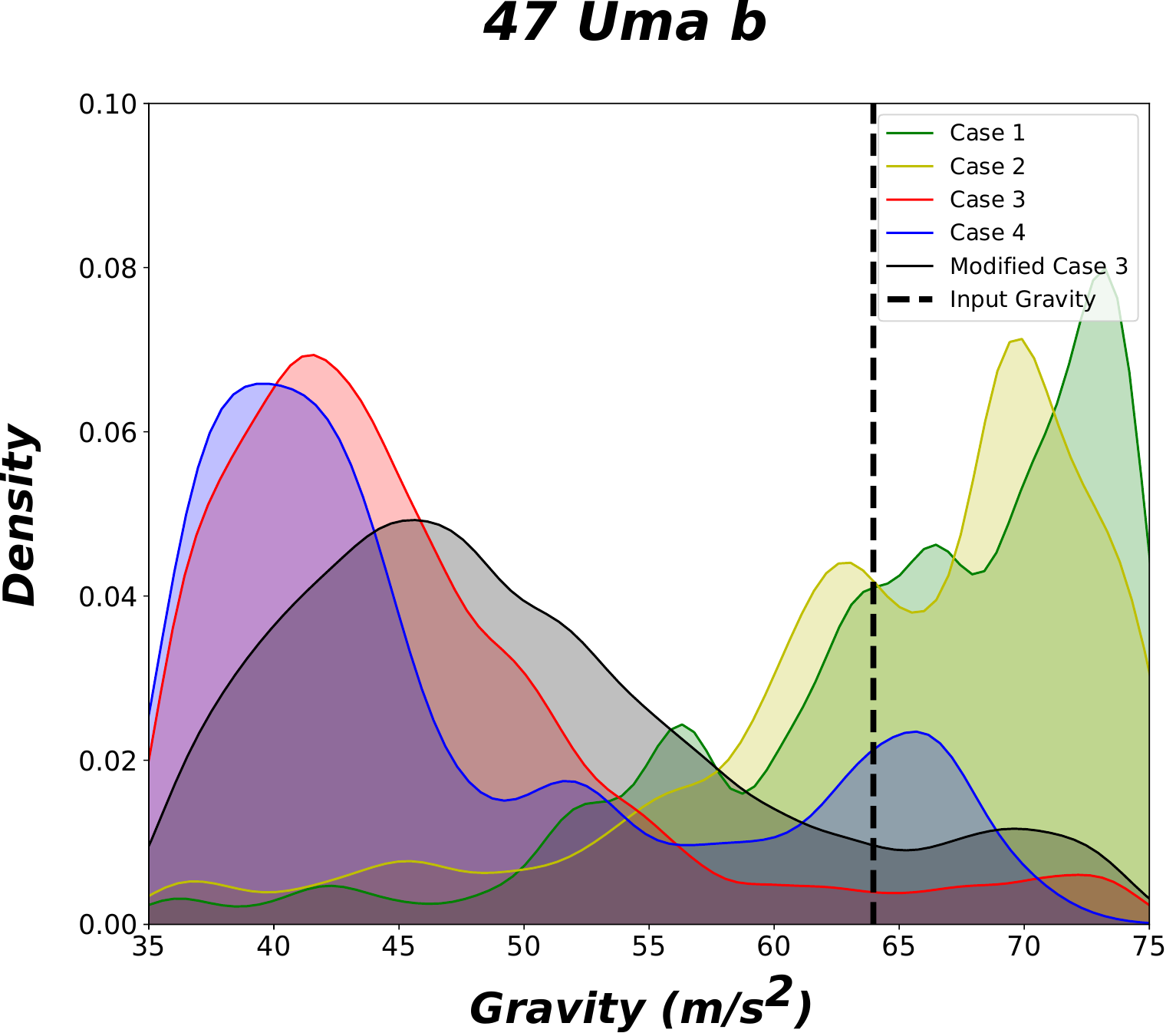}
  
  \caption{Comparison of the retrieved gravity posteriors for the four cases and modified Case 3 on {\it 47 Uma b}. The black dashed line depicts the true gravity of the planet. The green shaded curve depicts the posterior for Case 1 while the blue shaded curve depicts the posterior for Case 2. The posterior for Case 3 is depicted by the red shaded curve, modified Case 3 is depicted by the black shaded curve and that for Case 4 is depicted with the blue shaded curve. {\bf Main Point}- No parameterization can accurately improve the gravity beyond the $\pm$50\% \textit{a prior} value.}
\label{fig:fig7}
\end{figure}

In order to determine the robustness of our results with respect to imprecise gravity measurements, we allow the gravity of our planet cases to vary by  $\pm$50\% of the assumed mass. With this level of uncertainty we then can explore whether or not the knowledge of the mass can be improved with the observed 0.3-1$\mu$m reflected light spectroscopy. Figure \ref{fig:fig7} shows the posteriors for the retrieved gravity for the five parameterizations explored for {\it 47 Uma b} in \S \ref{sec:47umab}. 

Case 1 and Case 2 retrieve gravity posteriors which clearly favor higher gravity values whereas Case 3 (no water depletion), Case 3 (with water depletion) and Case 4 show posteriors peaking toward lower values of gravity (beyond 2$\sigma$) compared to the true input gravity shown by the dotted line in Figure \ref{fig:fig7}. %Therefore, none of the parameterizations reliably improve the gravity constraint. 
From this we draw two conclusions: 1) none of the parameterizations here could reliably retrieve gravity with the spectral resolution and SNR of reflected spectra used in this work , 2) even with $\pm$50\% inaccuracy in gravity cloud structures and abundances of molecules can be inferred from the reflected spectra of cool giants with proper choice of cloud parameterizations. Further analysis, beyond the scope of this analysis, would need to be performed to determine if these conclusions were: 1) robust against gravity constraints that were larger than $\pm$50\%, and 2) robust against unconstrained phase angle (see further discussion on phase \S \ref{sec:phase}).

%%%%%%%%%%% THIS FIGURE HAS BEEN UPDATED WITH NEW RUNS

\subsubsection{Effect of Assumed Signal-to-Noise Ratio}

Throughout the analysis we fixed SNR. We determined that for a planet like {\it 47 Uma b} the retrieved results were highly dependent on the complexity of the used retrieval model and the overall parameterization. In order to determine the robustness of this result with respect to the assumed SNR, we degrade the SNR to see whether this complexity dependence still holds for a simulated spectra of {\it 47 Uma b} with a lower SNR of 5.

\begin{figure*}
  \centering
  \includegraphics[width=0.45\textwidth]{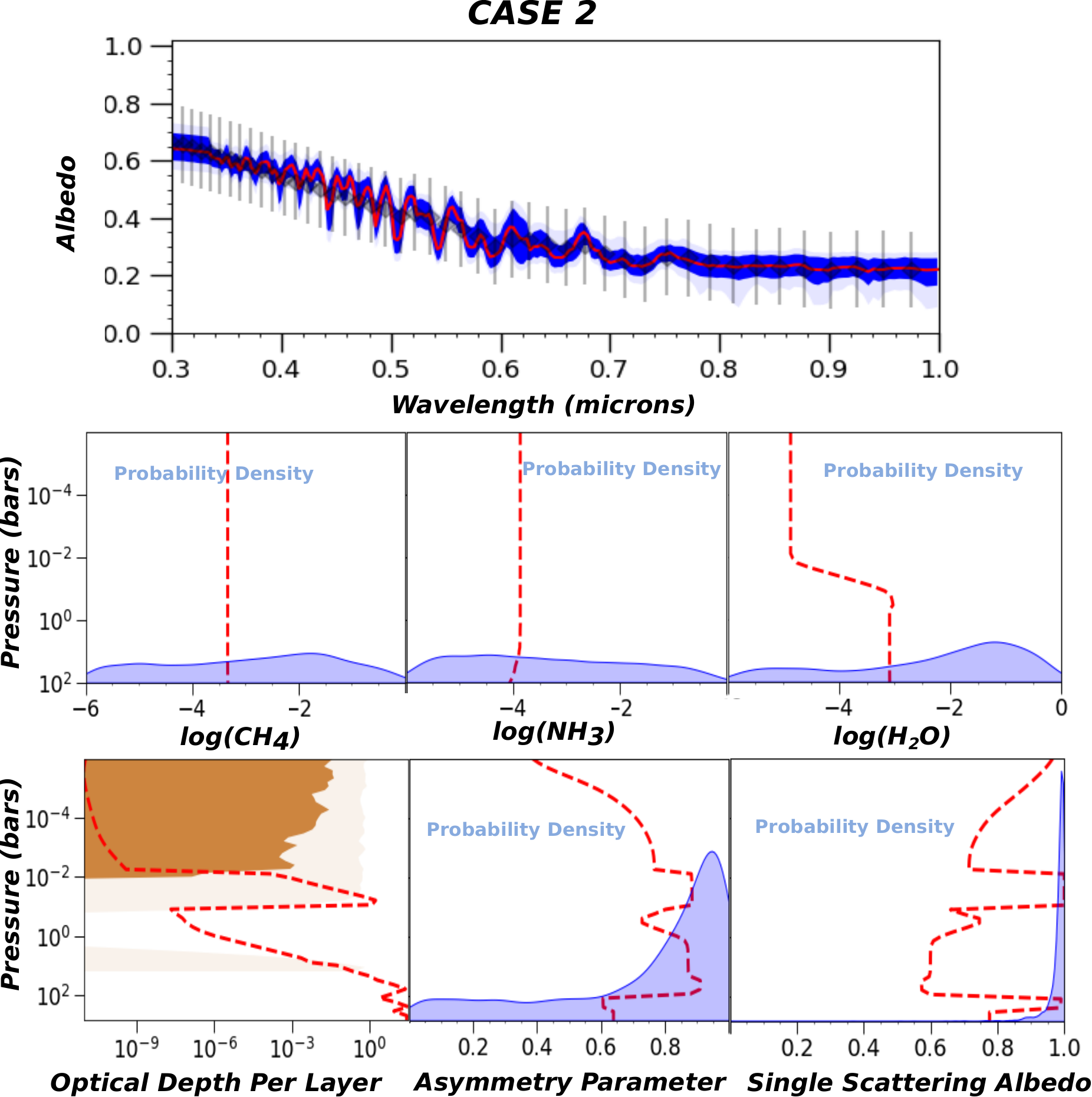}
  \includegraphics[width=0.45\textwidth]{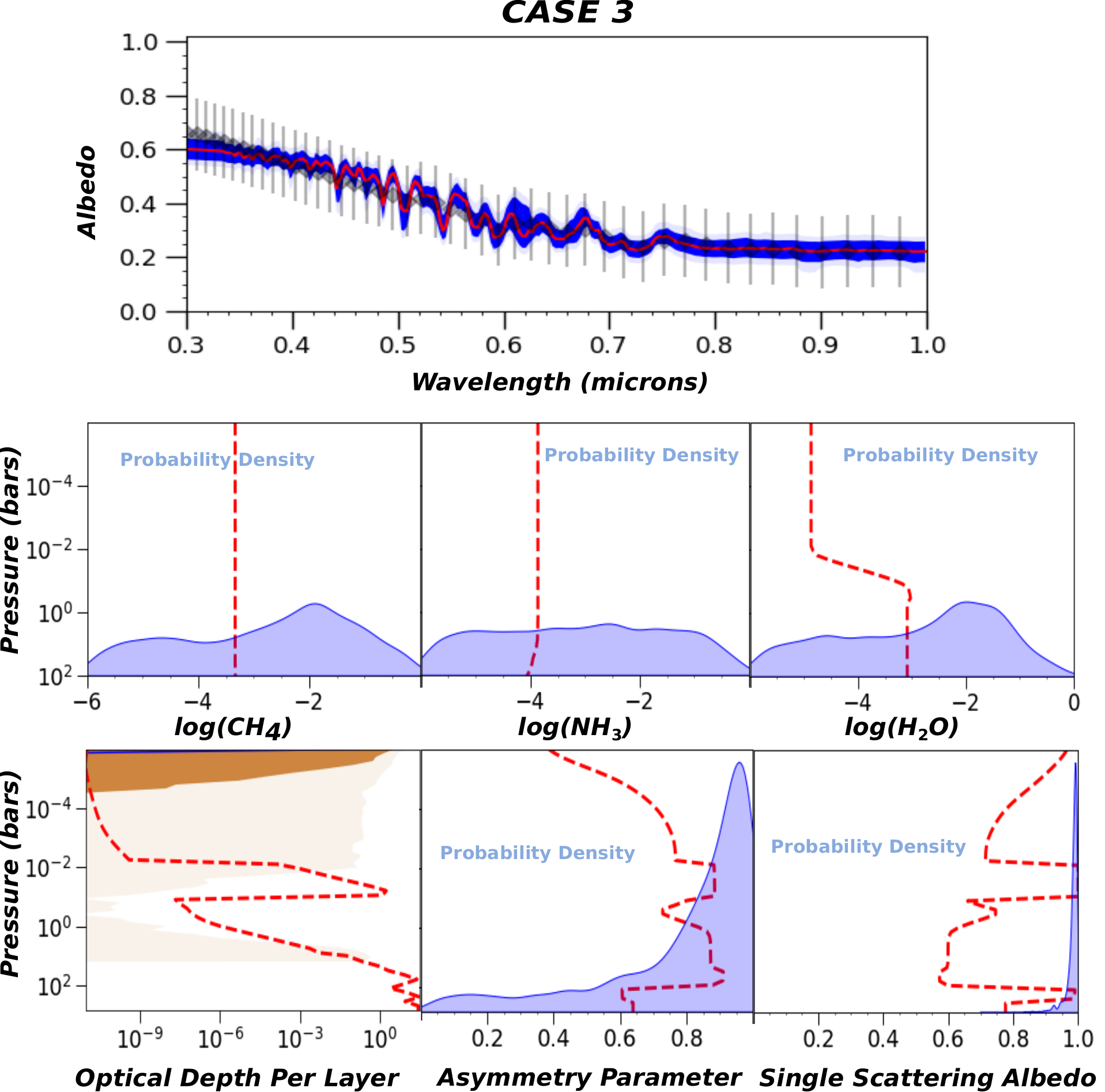}
  
  \caption{Comparison of the retrieved atmospheric properties of {\it 47 Uma b} on a spectra with SNR 5. The left panel shows the retrieval results using Case 2 parameterization while the right column shows retrieval results with Case 3 parameterizations. The top row shows the comparison of the observed spectra in black diamonds with the median (red), 1$\sigma$ (dark blue) and 2$\sigma$ (light blue) regions of the retrieved spectra with each parameterization. The middle row shows the retrieved mixing ratios of CH$_4$, NH$_3$ and H$_2$O from left to right in both the columns with the true input profiles shown in red. The last row shows the retrieved cloud properties -- optical depth per layer, asymmetry parameter and single scattering albedo from left to right in both the columns with the true input profiles shown in red. {\bf Main Point}- At low SNR of 5 neither Case 2 and Case 3 are able to accurately or precisely retrieve molecular abundances.}
\label{fig:case2_snr5}
\end{figure*}

Figure \ref{fig:case2_snr5} shows the retrieval results on a spectra with SNR=5 for Case 2 and Case 3 parameterizations. At lower SNR, retrievals produced with Case 1 and Case 2 parameterizations are able to fit the observed spectra. This is an intuitive result as the extra absorption features, which were seen at SNR=20, are now buried within the systematic error bars of the simulated spectra. Similar to previous results of \citet{Lupu_2016} and \citet{Feng_2018}, we find that at such low SNR, none of the cases result in precise or accurate constraints of molecular abundances.  According to the Bayes Factor, Case 2 rules out Case 3 very weakly with this quality of data. %Similarly, the cloud structure or optical properties are not accurately or precisely constrained by either of the two cases.  

\begin{figure}
  \centering
  \includegraphics[width=0.5\textwidth]{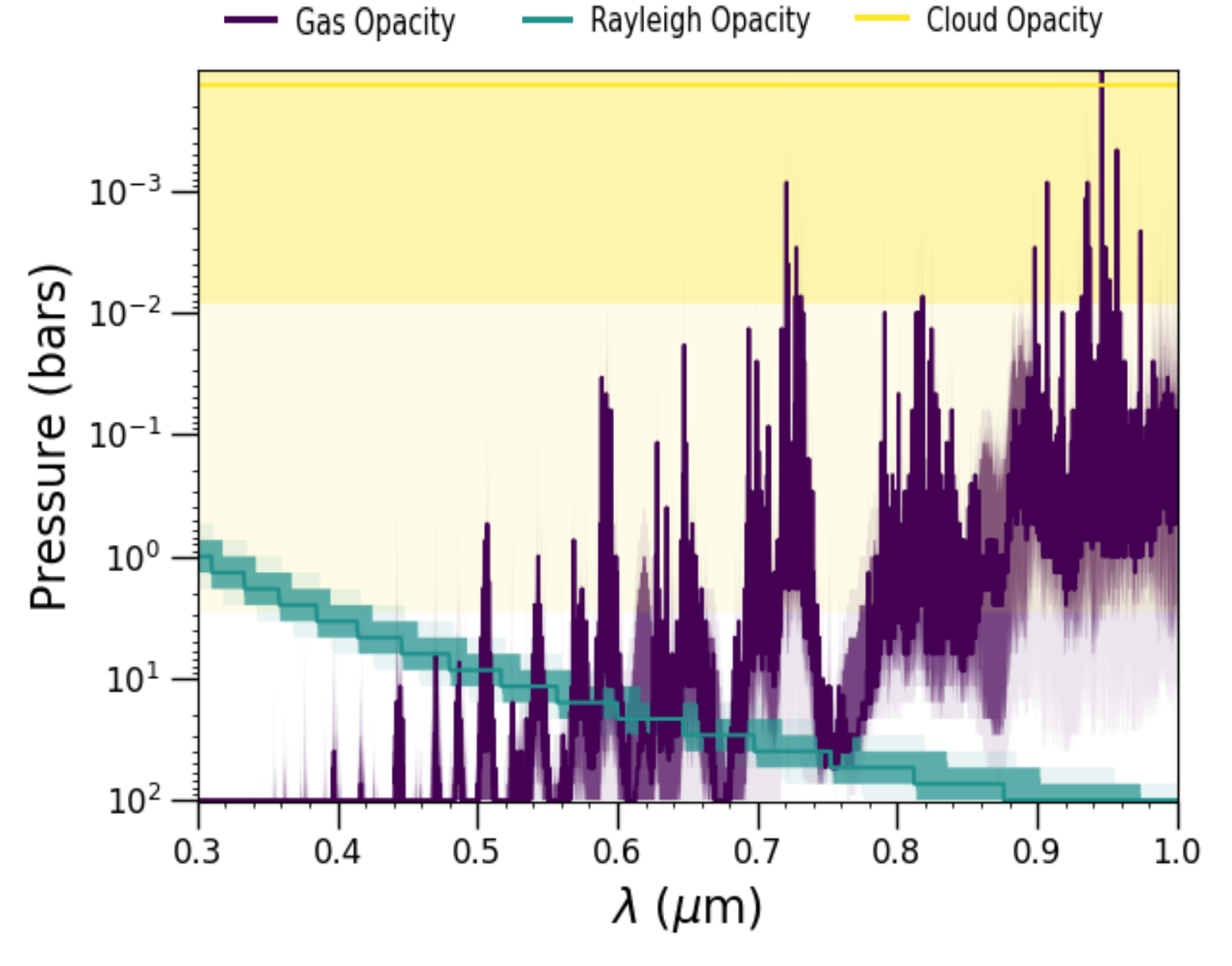}
  
  \caption{Photon attenuation depth map corresponding to an optical depth of 0.5 retrieved using Case 2 from the SNR=5 spectra of {\it 47 Uma b}. The attenuation pressure levels are divided into Rayleigh, gas and cloud opacity. The dark shaded yellow, purple and green regions depict 1$\sigma$ regions around the retrieved cloud, gas and Rayleigh opacity levels, respectively. The light shaded region of the same colors depict the 2$\sigma$ intervals.}
\label{fig:case2_snr5_phot}
\end{figure}

Despite not being able to constrain molecular abundances directly, we can make inferences as to where the cloud deck is relative to the molecular and Rayleigh opacity levels based on the retrieved photon attenuation map. We demonstrate this with the retrieved photon attenuation map for Case 2 shown in Figure \ref{fig:case2_snr5_phot}. Comparing this retrieved photon attenuation map with that shown in Figure \ref{fig:figphot} shows that the cloud optical depth level can be retrieved within 2$\sigma$ of the `true' opacity levels with the Case 2 parameterization. The Rayleigh opacity levels and the gas opacity levels are also retrieved within 1$\sigma$ of their respective `true' optical depth. This estimate of the cloud base pressure level from a SNR=5 albedo spectra can roughly and indirectly inform temperature-pressure structure of the atmosphere. This is because the location of the cloud deck is predicated on the region where the temperature becomes cool enough to condense a respective species. Therefore, by combining the expected equilibrium temperature of the planet, and the retrieved cloud deck, zeroth order inferences can be made about the potential temperature regime of the atmosphere. 

This analysis demonstrates that at lower SNR ($\sim$ 5) observations, single cloud deck parameterizations are preferred when performing retrievals (even when two cloud decks are present). We also verify that all the models fail to capture any of the atmospheric properties like molecular abundances directly with this data quality. This remains true even at higher resolutions of 140. However, inspection of the retrieved photon attenuation maps for lower SNR observations can be informative regarding the positions of the cloud opacity and therefore, the planet's climate.   

%%%%%%%%%%% THIS FIGURE HAS BEEN UPDATED WITH NEW RUNS

%\subsection{Retrieving Well-Mixed Chemistry vs. Depleted Chemistry due to Cloud Formation}

%Condensation of molecules leads to their depletion above the cloud base. This effect is seen clearly among two of our three target planets in Figure \ref{fig:figchem}. We deliberately begin our retrievals under the well-mixed assumption in order to determine if additional complexity is warranted. However, as seen in Figure \ref{fig:fig5}, the retrieved Case 3 H$_2$O mixing ratio posterior distribution of {\it 47 Uma b} was biased towards lower H$_2$O abundance. Hence we added another level of complexity for retrieving a H$_2$O profile depleted above the top cloud base along with Case 3 for {\it 47 Uma b}. This led to improved constraints on the cloud structure of the planet with Case 3. This indicates that at SNR$\sim$20 and R$\sim$40 spectra are sensitive to depleted abundances of major opacity sources above the cloud deck. At lower SNR=5, though, depletion of abundances is no longer necessary to consider. This is because the constrained abundances are worse when retrieved with a retrieval model having Case 3 like complexity and this sensitivity can significantly improve the retrieved cloud structure as well.

\subsection{Validity of Wavelength Independent Cloud Properties in our Retrieval Models}

We ignore any wavelength dependence while parametrizing the cloud optical properties for all of our retrieval models. However, when modeling our simulated data we consider full wavelength dependent cloud optical properties from {\texttt{Virga}}. This begs the question of whether or not additional wavelength-dependent cloud complexity would be needed for our SNR/R and wavelength range parameter space. For the cases considered here, though, this additional complexity is not necessary because of the: 1) wavelength region explored, and 2) optical properties of the cloud species explored.

The cloud optical properties of {\it eps Eri b} and {
\it 47 Uma b} show negligible wavelength dependence. This is because the optical properties of H$_2$O and NH$_3$ are not strongly wavelength dependent within the wavelength range of our focus (0.3-1 microns). Therefore, this assumption is strongly dependent on the particular species explored. For planets that are relatively hotter than {\it eps Eri b} and {\it 47 Uma b}, condensation of sulfur-based species such as MnS, Na$_2$S and ZnS, may occur. These species show strong wavelength dependence in their scattering properties \citep[e.g.][]{querry1987optical}. Although our hottest target {\it HD 62509 b} has a cloud deck dominated by sulfur species (Na$_2$S), the overall opacity is dominated by Rayleigh and molecular contribution. That is because the cloud deck of {\it HD 62509 b} is far too low in altitude (high in pressure). Therefore the necessity of wavelength dependent properties is not warranted.

Lastly, in addition to condensates, hazes can add an additional wavelength dependence. In particular, sulfur hazes, which strongly absorb light toward 0.3$\mu$m, can create positive-sloped spectra \citep{gao2017sulfur} that would require the consideration of a wavelength-dependent cloud retrieval. A Jupiter-like, wavelength-dependent haze has also been retrieved by using a simple parameterization \citep{lacy2018wfirst}. These specific cases are beyond the scope of this analysis, but could add an additional level of complexity to the parameterizations explored here. 

\subsection{Additional Uncertainty Caused By Unknown Phase Angle} \label{sec:phase}

Throughout our analysis the phase angle of our target planets have been kept to zero. Phase, however, changes the albedo spectrum of a planet significantly because the scattering properties of the atmosphere are phase dependent. Therefore, an unknown phase can lead to additional uncertainty in the retrieval analysis. This effect has been the subject of previous exploration. In particular, there exists a known degeneracy between the phase angle and the radius of the planet \citep{nayak2017atmospheric}. \citet{nayak2017atmospheric} showed that an unknown phase angle does not lead to a significant change in the accuracy of retrieved molecular abundances and cloud structure, when compared to the case of known phase angle. However, when retrieving on contrast (relative planet-to-star ratio) as opposed to albedo, the unknown phase angle does introduce significant uncertainty in the radius retrieval of the planet compared to the case where phase angle is known \citet{nayak2017atmospheric}.

To test the sensitivity to phase angle in this analysis, we performed a retrieval on the albedo spectra of {\it 47 Uma b} simulated at a phase angle of 90$^{\circ}$ with our Case 3 retrieval model. We assume a uniform prior for the phase angle between 60$^{\circ}$ and 120$^{\circ}$. Similar to \citet{nayak2017atmospheric}, we find that the precision and accuracy of the retrieved molecular abundances are unchanged relative to the case of zero phase, within 1$\sigma$. We do find significant bi-modality in the retrieved cloud solution of the 1) pressure level of the optically thick (high pressure) cloud deck, 2) the gravity, and 3) the phase angle. Instead of retrieving a single peaked posterior at 30~bars, a double peaked posterior solution of 30~bars and 0.3~bars is retrieved. Therefore, the combination of unknown gravity and phase angle will impede accurate and precise determination of cloud properties. However, given the solution is strongly bimodal, inferences could be made regarding the most likely physical scenario.

\section{Conclusions and Future Work}\label{sec:conclusion}
We have performed retrievals on reflected light albedo spectra for three high priority cool giant targets for future space-based optical high-contrast imaging and spectroscopic missions like {\it HabEx} and {\it LUVOIR}. We have chosen planets with three different estimated effective temperatures of 135 K, 217 K and 533 K. This wide range of cool giant effective temperatures helps to explore retrievals of various possible cloud structure scenarios for cool giants. Albedo spectra for these planets were calculated using the spectroscopy modeling code -- \texttt{PICASO} and robust cloud calculation model -- \texttt{Virga}. We used the modeled albedo spectra to simulate mock observation spectra with a constant spectral resolution of 40 and a SNR of 20. Here we briefly discuss the key aspects and results of our retrieval analysis.
\begin{enumerate}

\item Requisite cloud complexity is highly sensitive to the relative position of the molecular, cloud, and Rayleigh opacity. The additional complexity of a second cloud deck, for example, is only favored (according to Bayes factor analysis) when the region of highest molecular opacity contribution is between the two cloud decks. Otherwise, the atmosphere can be simply parameterized with a larger, single deck. 

\item Box model parameterizations for cloud opacity result in abundance measurements that are largely over-estimated. Therefore, exponential cloud opacity parameterizations (e.g. at least Case 2) should be used instead, even at low SNR$\sim$5 observations. 

\item Although single scattering and asymmetry of the cloud deck changes with altitude, a 2-valued model for these scattering properties never retrieves a more accurate solution than single-valued models for these scattering properties (i.e. Case 3 retrieves the scattering properties more accurately than Case 4). This conclusion, however, might only apply to the planets explored here (dominated by NH$_3$ and H$_2$O clouds). Planets with two cloud decks composed of condensates or hazes with drastically different optical properties might warrant additional altitude-based complexity.

%\item We detect sensitivity to the H$_2$O depletion above the cloud deck due to water condensation for the case of {\it 47 Uma b}. 
\item Allowing for an altitude-dependence in the H$_2$O mixing ratio profile in order to detect H$_2$O depletion above the cloud deck due to water condensation for the case of {\it 47 Uma b}, slightly improves the precision and accuracy of the abundances. However, this seemingly ``better'' solution was weakly rejected over an identical retrieval without altitude dependence. Therefore, although the fit appears better (i.e. increased precision and accuracy with respect to the 1$\sigma$ constraint interval), the additional complexity is not statistically favored for this data quality. 

\item We find that even at very low SNR=5, low R=40 (0.3-1$\mu$m), inferences can be made with respect to the position of the cloud deck without attaining accurate information regarding the abundances of molecular species. In accordance with other works \citep{Lupu_2016,nayak17,Hu_19} we are unable to attain precisely constrained molecular abundances with this data quality. However, we are able to retrieve a photon attenuation map of the expected opacity contribution of rayleigh scattering, cloud scattering, and molecular absorption. This gives a limit as to the position of the bottom of the cloud deck. This suggests that very coarse, zeroth-order, temperature information could be attained by combining the equilibrium temperature of the planet, with knowledge of condensation curves. 

\item Lastly, we show that the cloud structure and molecular mixing ratios of the planets can be accurately and precisely retrieved with a $\pm$50\% uncertainty in the gravity of the planets. However, it is not possible to improve the gravity constraint beyond this value.

\end{enumerate}

Returning to our initial questions posed: 1) %\texttt{PICASO} can be used to perform retrievals of reflected light observations. All of the parameterizations used in this analysis can be found within \texttt{PICASO}. 2) 
Users' choice of cloud and atmospheric parameterization strongly effect the precision and accuracy of the resultant abundances and cloud structure. 2) The specific location of the cloud deck, with respect to the location of the optically thick molecular opacity, dictates whether or not accurate cloud structure information can be retrieved though this information will not be known \textit{a prior}. 3) Precise gravity information is very difficult to retrieve with the quality of simulated data used here, but atmospheric characterization with reflected light is possible even with large uncertainties in planet gravity. 4) Lastly, even low SNR=5, low R=40 reflected light spectroscopy from 0.3-1$\mu$m can give insights into the cloud deck position of the planet. 

\section{Acknowledgments}
SM would like to thank the S.~N.~Bose Scholar's program by Indo-US Science and Technology Forum (IUSSTF) for funding his visit to the Department of Astronomy and Astrophysics, UC Santa Cruz through the S.~N.~Bose Scholarship. MM acknowledges the support of the Nancy Grace Roman Science Investigation Team program. The authors would like to thank the exoplanet group at UC Santa Cruz especially Jonathan Fortney for all the computational support and resources used in this work. The authors will also like to thank Ryan MacDonald for insightful discussions and the anonymous referee for their suggestions, which helped in improving the manuscript.
 
{\it Software:} PICASO \citep{batalha19}, Virga \citep{batalha20}, DYNESTY \citep{speagle20}, numba \citep{numba}, pandas \citep{mckinney2010data}, bokeh \citep{bokeh}, NumPy \citep{walt2011numpy}, IPython \citep{perez2007ipython}, Jupyter \citep{kluyver2016jupyter},PySynphot \citep{pysynphot2013}, sqlite3, matplotlib \citep{Hunter:2007}, PyMieScatt \citep{sumlin18retrieving}

\bibliography{sample63}{}
\bibliographystyle{apj}



\end{document}